\documentclass[preprint,3p]{elsarticle}

\usepackage{graphicx}
\usepackage{amsmath}
\usepackage{array}
\usepackage{amssymb}
\usepackage{multirow}
\usepackage{subfig}

\journal{Journal of Computational Physics}

\newcommand{\Jn}{J_n}

\newcommand{\Hn}{H^{(1)}_n}

\newcommand{\In}{I_n}
\newcommand{\Ind}{I'_n}
\newcommand{\Kn}{K_n}
\newcommand{\Knd}{K'_n}

\newcommand{\hIn}{\hat{I}_n}
\newcommand{\hInd}{\hat{I}'_n}
\newcommand{\hKn}{\hat{K}_n}
\newcommand{\hKnd}{\hat{K}'_n}

\newcommand{\tR}{\widetilde{R}}
\newcommand{\hR}{\hat{R}}
\newcommand{\htR}{\hat{\widetilde{R}}}
\newcommand{\tT}{\widetilde{T}}
\newcommand{\hT}{\hat{T}}
\newcommand{\htT}{\hat{\widetilde{T}}}

\newcommand{\rh}{\boldsymbol{\rho}}
\newcommand{\rhp}{\boldsymbol{\rho'}}

\newcommand{\suma}{\sum_{n=-\infty}^{\infty}}
\newcommand{\sumb}{\sum_{n=1}^{\infty}}

\newcommand{\intzi}{\int_{0}^{\infty}}

\newcommand{\iu}{\mathrm{i}}
\newcommand{\lam}{\lambda}

\begin{document}

\begin{frontmatter}
\title{Computation of potentials from current electrodes in cylindrically stratified media: A stable, rescaled semi-analytical formulation}

\author[rvt]{H. Moon\corref{cor1}}
\ead{moon.173@osu.edu}
\author[rvt]{F. L. Teixeira}
\ead{teixeira@ece.osu.edu}
\author[rvt2]{B. Donderici}
\ead{burkay.donderici@halliburton.com}

\cortext[cor1]{Corresponding author}
\address[rvt]{ElectroScience Laboratory, The Ohio State University, Columbus, OH 43212, USA}
\address[rvt2]{Sensor Physics \& Technology, Halliburton Energy Services, Houston, TX 77032, USA}

\begin{abstract}
We present an efficient and robust semi-analytical formulation to compute the electric potential due to arbitrary-located point electrodes in three-dimensional cylindrically stratified media, where the radial thickness and the medium resistivity of each cylindrical layer can vary by many orders of magnitude. A basic roadblock for robust potential computations in such scenarios is the
poor scaling of modified-Bessel functions used for computation of the semi-analytical solution, for extreme arguments and/or orders.
To accommodate this, we construct a set of rescaled versions of modified-Bessel functions, which avoids underflows and overflows in finite precision arithmetic, and minimizes round-off errors. In addition, several extrapolation methods are applied and compared to expedite the numerical evaluation of the (otherwise slowly convergent) associated Sommerfeld-type integrals. The proposed algorithm is verified in a number of scenarios relevant to geophysical exploration, but the general formulation presented is also applicable
to other problems governed by Poisson equation such as Newtonian gravity, heat flow, and potential flow in fluid mechanics, involving cylindrically stratified environments.

\end{abstract}


\end{frontmatter}

\section{Introduction}
\label{sec.1.intro}
Resistivity logging is extensively used for detecting, characterizing, and analyzing hydrocarbon-bearing zones in the subsurface
earth~\cite{Wait78:Analysis, Moran79:Effects, Gianzero82:Integral, Drahos84:Electrical, Lovell90:Effect, Lovell:thesis:Finite, Zhang02:Real, Doestch10:Borehole}.
This sensing modality employs electrode-type devices mounted on a mandrel that inject electric currents into the surrounding earth formation~\cite{Ellis:Well, Telford:Applied}. The ensuing electric potential is then measured at different locations to provide estimates for the surrounding resistivity.
Many numerical techniques such as finite-differences, finite elements, numerical mode-matching, and finite volumes method can be used to model the response of resistivity logging tools
~\cite{Liu02:Electromagnetic,Hue05:Three, Lee12:Numerical,Sasaki94:3D, Pardo06:Two, Pardo06:Simulation, Nam10:Simulation, Ren10:3D,Novo07:Finite, Novo10:Three,Liu94:Modeling, Fan00:3D, Hue07:Numerical}.
Brute-force techniques are rather versatile and applicable to arbitrary resistivity distributions; however, at the same time, this precludes optimality in particular cases of special interest, such as when resistivity logging environment can be represented as a cylindrically stratified medium~\cite{Chew:Waves}. Depending on the implementation, brute-force techniques may have difficulties handling extreme sharp discontinuities in the coefficients, as is the case for the resistivity parameter for the physical scenario considered here, which can change by many orders of magnitude across adjacent layers.

In this paper, a robust semi-analytical formulation for computing the electric potential due to arbitrary-located point electrodes in three-dimensional cylindrically stratified media is proposed. The present formulation is based on a series expansion in terms of azimuth Fourier modes and a spectral integral over the vertical wavenumber along the axial direction. The resulting problem in terms of the radial variable yields a set of modified Bessel equations.
The present formulation removes roadblocks for numerical computations associated with the poor scaling of modified-Bessel functions for very small and/or very large arguments and/or orders~\cite{Smythe:Static,Wait:Geo,Carley13:Bessel}. This is done by constructing a set of rescaled, modified-Bessel functions that can be stably evaluated under double-precision arithmetic,
akin to what has been done in the past for ordinary (non-modified) Bessel functions~\cite{Moon14:Stable} .
The present formulation also carefully manipulates the analytical formulae for the potential in such media to yield a set of integrand expressions can be computed in a robust manner under double-precision for a wide range of layer thicknesses, layer resistivities, and source and observation point separations.
Finally, a number of acceleration techniques are implemented and compared to effect the efficient numerical integration of the Sommerfeld-type (spectral) integrals, which otherwise suffer from slow convergence.
The proposed algorithm is verified in a number of practical scenarios relevant to geophysical exploration. 
More generally, the mathematical setting here corresponds to the classical problem of obtaining the Green's function for the steady diffusion equation (Poisson problem) with discontinuous coefficients in a separable geometry.  As such, the general formalism presented here is also applicable to other problems governed by Poisson equation such as Newtonian gravity, heat flow, elasticity, neutron transport, and potential flows in fluid mechanics, in cylindrically stratified geometries.

\section{Formulation}
\label{sec.2.formulation}
\subsection{Electric potential in homogeneous media}
\label{sec.2.1}
In a homogeneous medium, the electric potential $\psi$ from a current electrode at the origin writes as~\cite{Wait:Geo}
\begin{flalign}
\psi
=\frac{\mathcal{I}}{4\pi\sigma\sqrt{\rho^2+z^2}}
=\frac{\mathcal{I}}{2\pi^2\sigma}\intzi K_0(\lam\rho)\cos(\lam z)d\lam, \label{psi.homo}
\end{flalign}
where $\mathcal I$ is the electric current flowing into the medium from the electrode, $\sigma$ is the conductivity of the medium, and $K_0(\cdot)$ is the modified-Bessel function of the second kind of the zeroth order. For the second equality, the complete Lipschitz-Hankel integral \cite{Watson:Bessel} is employed. When the source is off the origin, higher order azimuthal modes appear. Using the addition theorem for $K_0$, \eqref{psi.homo} is modified to
\begin{flalign}
\psi&=\frac{\mathcal{I}}{2\pi^2\sigma}\intzi K_0(\lam|\rh-\rhp|)\cos(\lam (z-z'))d\lam \notag\\
	&=\frac{\mathcal{I}}{2\pi^2\sigma}\suma e^{\iu n(\phi-\phi')}\intzi
	\In(\lam\rho_<)\Kn(\lam\rho_>)\cos(\lam (z-z'))d\lam, \label{psi.off.origin}
\end{flalign}
in terms of modified-Bessel functions of both first, $I_n(\cdot)$ and second, $K_n(\cdot)$, kinds. In the above, primed coordinates ($\rho',\phi',z'$) represent the source location and unprimed coordinates ($\rho,\phi,z$) represent the observation point. Also, $\rho_<=\min(\rho,\rho')$ and $\rho_>=\max(\rho, \rho')$.

\subsection{Electric potential in cylindrically stratified media}
\label{sec.2.2}
\begin{figure}[h]
    \centering
    \includegraphics[width=2.0in]{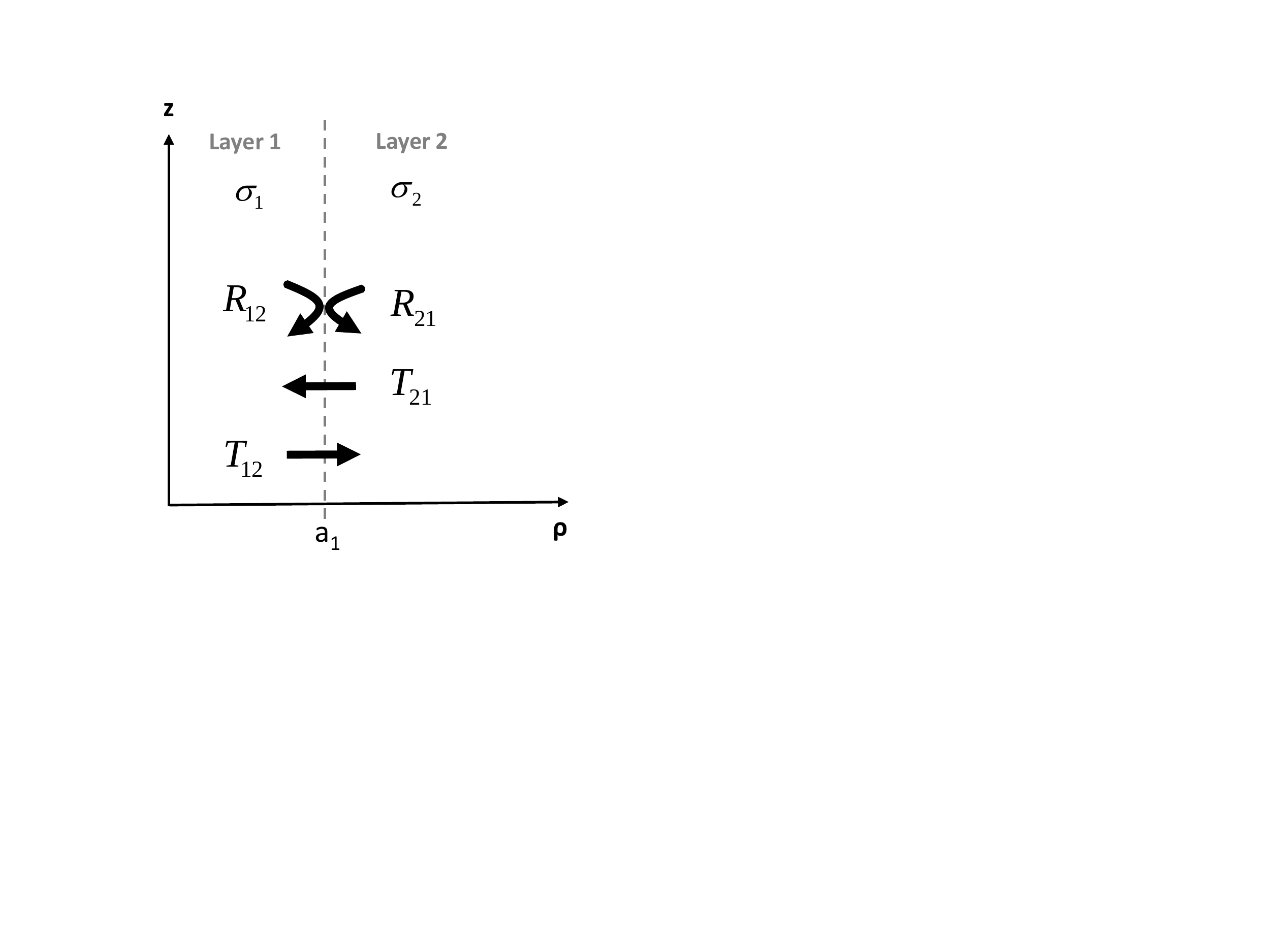}\\
    \caption{Schematic description of two layers with associated coefficients in the $\rho z$-plane.}
    \label{F.2layers}
\end{figure}

In a cylindrically stratified medium, boundary conditions at the interfaces need to be incorporated. Let us first consider the case with two distinct cylindrical layers, as depicted in Fig. \ref{F.2layers}.
When the source is embedded in layer 1, we denote it the {\it outgoing-potential} case. In this case, the primary potential $\psi^{p}$ is a function of $\Kn(\lam\rho)$ because $\In(\lam\rho)$ diverges as $\rho$ goes to infinity. On the other hand, when the source is embedded in layer 2, we denote it the {\it standing-potential} case and $\psi^{p}$ is a function of $\In(\lam\rho)$ instead of $\Kn(\lam\rho)$ because $\Kn(\lam\rho)$ diverges when $\rho$ goes to zero.
For the outgoing-potential case, the $n$-th harmonic with $e^{\iu n(\phi-\phi')}$ dependence in layer 1 and layer 2 can be expressed, resp., as
\begin{subequations}
\begin{flalign}
\psi_1&=\left[\Kn(\lam\rho)+R_{12}\In(\lam\rho)\right]A_0, \label{2layers.out.psi1} \\
\psi_2&=T_{12}\Kn(\lam\rho)A_0, \label{2layers.out.psi2}
\end{flalign}
\end{subequations}
where $R_{12}$ and $T_{12}$ are the (local) reflection and transmission coefficients at the boundary $a_1$, and $A_0$ is an arbitrary amplitude of the primary potential. Applying the boundary conditions~\cite{Wait:Geo} at the interface, we obtain
\begin{subequations}
\begin{flalign}
R_{12}&=\frac{(\sigma_2-\sigma_1)\Kn(\lam a_1)\Knd(\lam a_1)}
	{\sigma_1\Ind(\lam a_1)\Kn(\lam a_1) - \sigma_2\In(\lam a_1)\Knd(\lam a_1)}, \label{2layers.out.R12} \\
T_{12}&=\frac{\sigma_1}{\lam a_1 \left[\sigma_1\Ind(\lam a_1)\Kn(\lam a_1)
			- \sigma_2\In(\lam a_1)\Knd(\lam a_1)\right]}. \label{2layers.out.T12}
\end{flalign}
\end{subequations}
For the standing-potential case, we similarly have
\begin{subequations}
\begin{flalign}
\psi_1&=T_{21}\In(\lam\rho)B_0, \label{2layers.stand.psi1} \\
\psi_2&=\left[R_{21}\Kn(\lam\rho)+\In(\lam\rho)\right]B_0, \label{2layers.stand.psi2}
\end{flalign}
\end{subequations}
and
\begin{subequations}
\begin{flalign}
R_{21}&=\frac{(\sigma_2-\sigma_1)\In(\lam a_1)\Ind(\lam a_1)}
	{\sigma_1\Ind(\lam a_1)\Kn(\lam a_1) - \sigma_2\In(\lam a_1)\Knd(\lam a_1)}, \label{2layers.stand.R21}\\
T_{21}&=\frac{\sigma_2}{\lam a_1 \left[\sigma_1\Ind(\lam a_1)\Kn(\lam a_1)
			- \sigma_2\In(\lam a_1)\Knd(\lam a_1)\right]}. \label{2layers.stand.T21}
\end{flalign}
\end{subequations}

\begin{figure}[t]
	\centering
	\subfloat[\label{F.out.R12}]{%
      \includegraphics[width=2.7in]{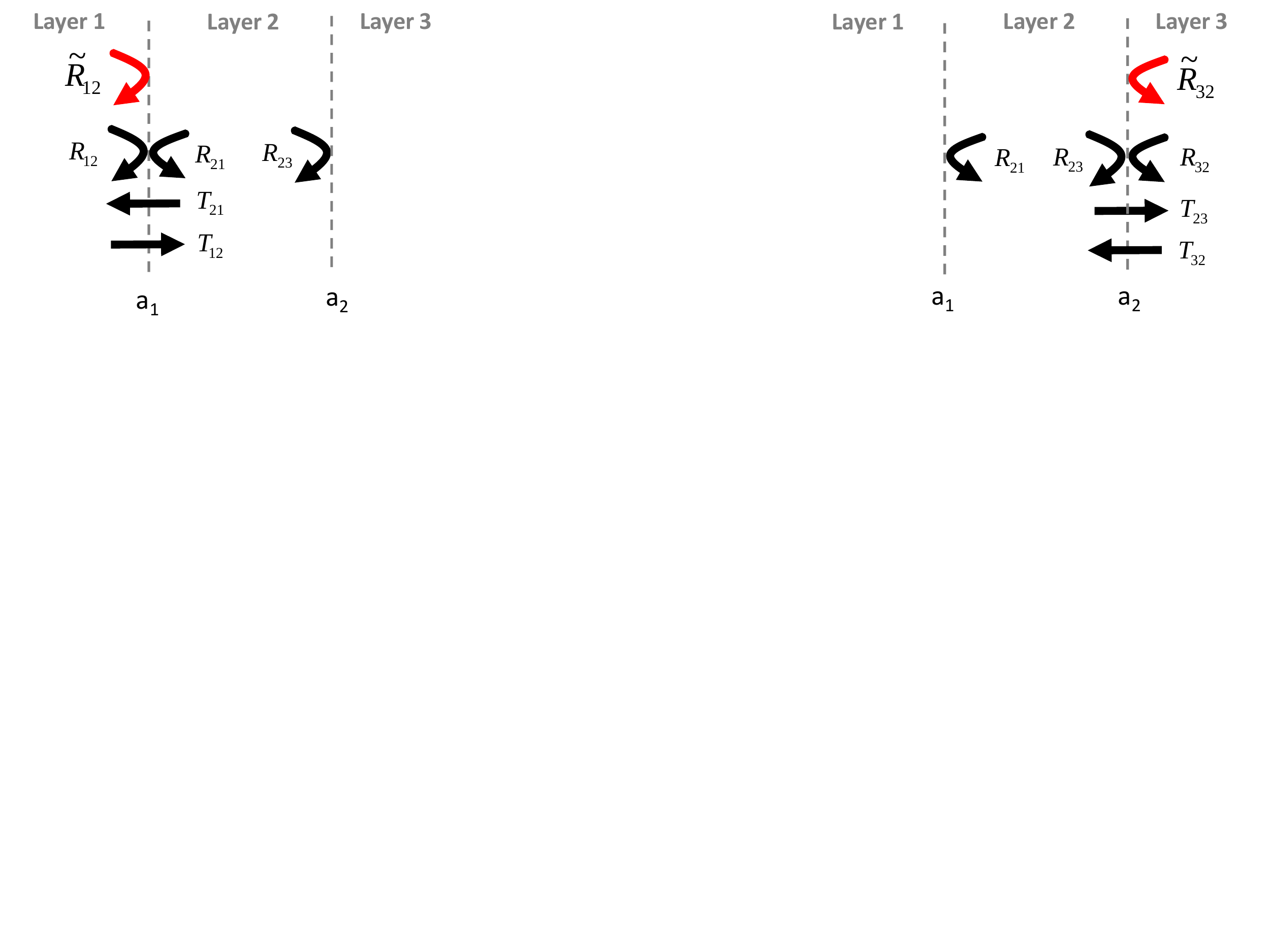}
    }
    \hspace{1.0cm}
    \subfloat[\label{F.out.S12}]{%
      \includegraphics[width=2.7in]{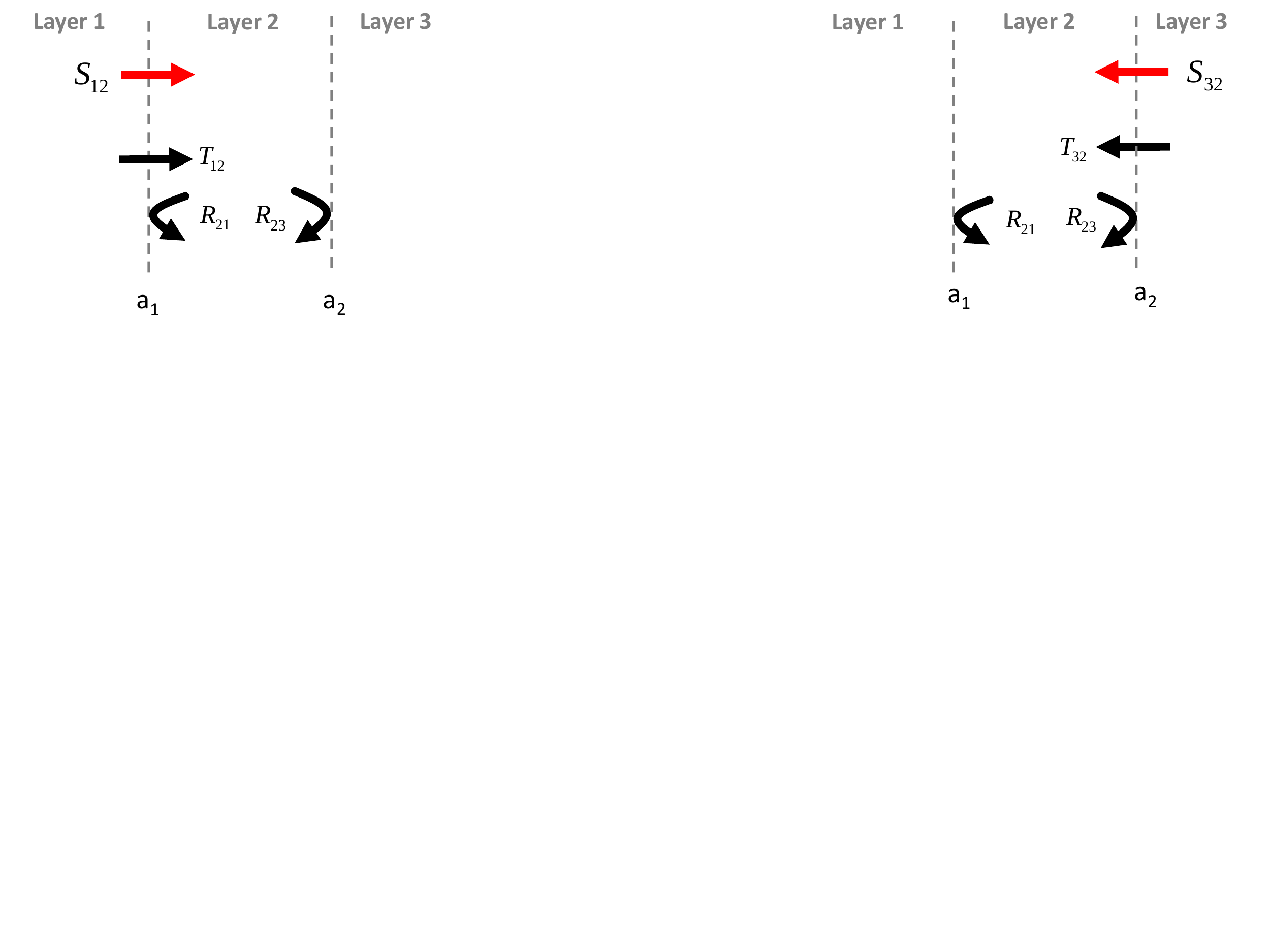}
    }
    \caption{Schematic description of three layers with associated coefficients for the outgoing-potential case: (a) $R_{12}$ and (b) $S_{12}$.}
    \label{F.3layers.outgoing}
\end{figure}

\begin{figure}[t]
    \centering
    \includegraphics[width=4.0in]{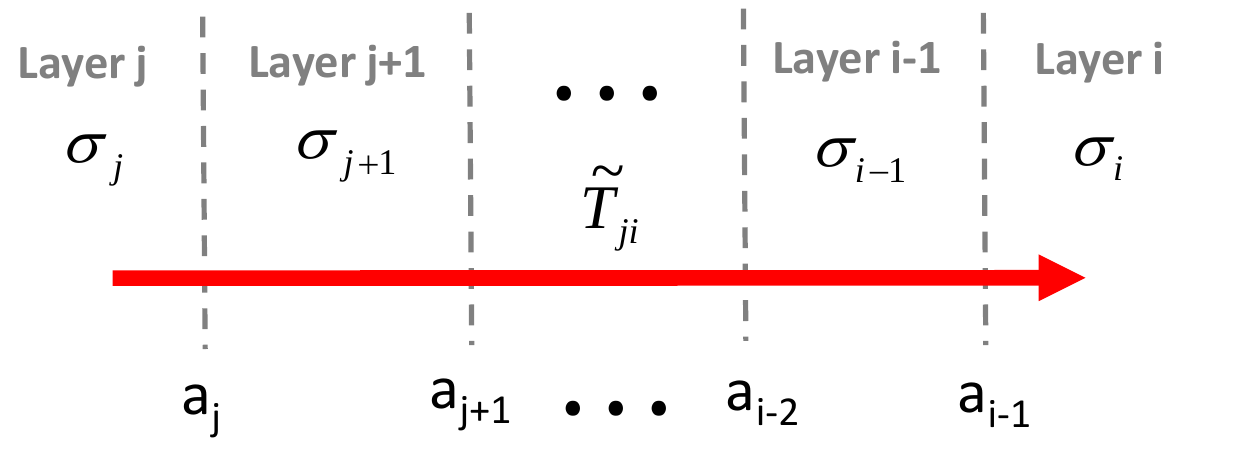}\\
    \caption{Generalized transmission coefficient for the outgoing-potential case.}
    \label{F.Tjicase1}
\end{figure}

When  more than two distinct layers are present, multiple reflections and transmissions occur. Therefore, generalized reflection and transmission coefficients should be determined. The procedure to obtain these coefficients is very similar to the one used for time-harmonic case in \cite{Chew:Waves} and will not be derived in detail here. Fig. \ref{F.3layers.outgoing} depicts the relevant coefficients to the outgoing-potential case in the medium consisting of three cylindrical layers. The potentials in the three layers can be expressed as
\begin{subequations}
\begin{flalign}
\psi_1&=\left[\Kn(\lam\rho)+\tR_{12}\In(\lam\rho)\right]A_1, \label{3layers.out.psi1} \\
\psi_2&=\left[\Kn(\lam\rho)+R_{23}\In(\lam\rho)\right]A_2, \label{3layers.out.psi2} \\
\psi_3&=T_{23}\Kn(\lam\rho)A_3. \label{3layers.out.psi3}
\end{flalign}
\end{subequations}
Applying proper constraint equations, we obtain
\begin{flalign}
\tR_{12}=R_{12}+T_{21}R_{23}(1-R_{21}R_{23})^{-1}T_{12}. \label{3layers.out.R12}
\end{flalign}
If another layer is added beyond layer 3, only $R_{23}$ needs to be replaced by $\tR_{23}$. Therefore, the generalized reflection coefficient in cylindrically stratified media for the outgoing-potential case is
\begin{flalign}
\tR_{i,i+1}=R_{i,i+1}+T_{i+1,i}\tR_{i+1,i+2}(1-R_{i+1,i}\tR_{i+1,i+2})^{-1}T_{i,i+1}.
	\label{multi.layers.out.Rij}
\end{flalign}
All amplitudes $A_i$'s as well as generalized reflection coefficients should be determined in order to obtain the potential everywhere. The relationship between two successive amplitudes in cylindrically stratified media can be written as
\begin{flalign}
A_{i+1}=T_{i,i+1}A_i + R_{i+1,i}R_{i+1,i+2} A_{i+1}. \label{multi.layers.out.A}
\end{flalign}
From \eqref{multi.layers.out.A}, a new coefficient denoted by $S$ is defined as
\begin{flalign}
S_{i,i+1}=(1-R_{i+1,i}\tR_{i+1,i+2})^{-1}T_{i,i+1} \label{multi.layers.out.S}
\end{flalign}
such that $A_{i+1}=S_{i,i+1}A_i$. The above coefficient can be regarded as a `local' transmission coefficient between two adjacent layers, as depicted in Fig. \ref{F.out.S12}.
Generalized transmission coefficient described in Fig. \ref{F.Tjicase1} can be defined using the $S$-coefficients \eqref{multi.layers.out.S} through
\begin{flalign}
\tT_{ji}=T_{i-1,i}S_{i-2,i-1}\cdots S_{j,j+1}
	    =T_{i-1,i}\prod_{k=j}^{i-2}S_{k,k+1}
        =T_{i-1,i}X_{j,i-1}. \label{out.Tji}
\end{flalign}
Note that $i>j$ for the outgoing-potential case. When $i=j+1$, $X_{j,i-1}=1$ in \eqref{out.Tji}.

\begin{figure}[t]
	\centering
	\subfloat[\label{F.stand.R32}]{%
      \includegraphics[width=2.7in]{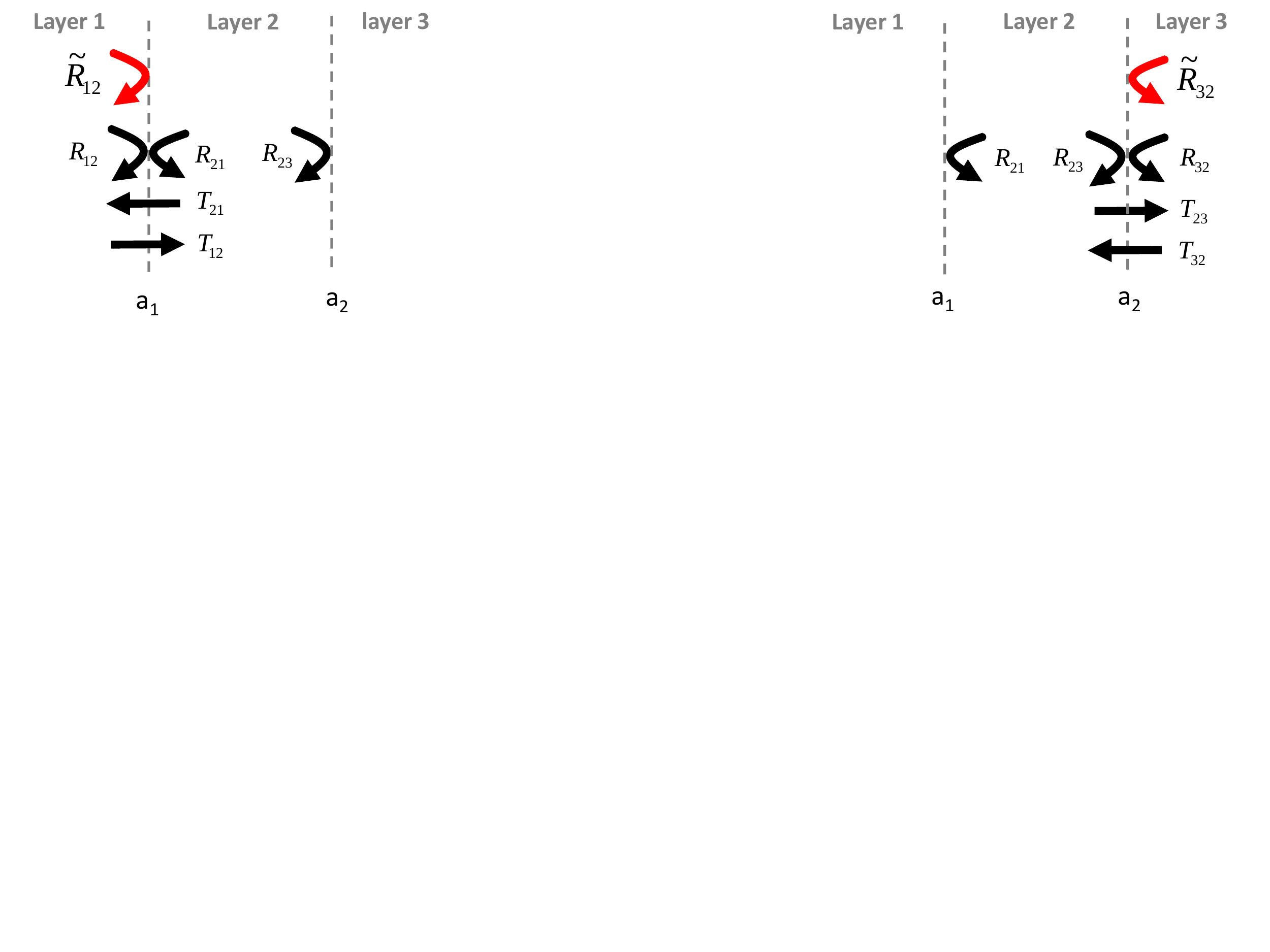}
    }
    \hspace{1.0cm}
    \subfloat[\label{F.stand.S32}]{%
      \includegraphics[width=2.6in]{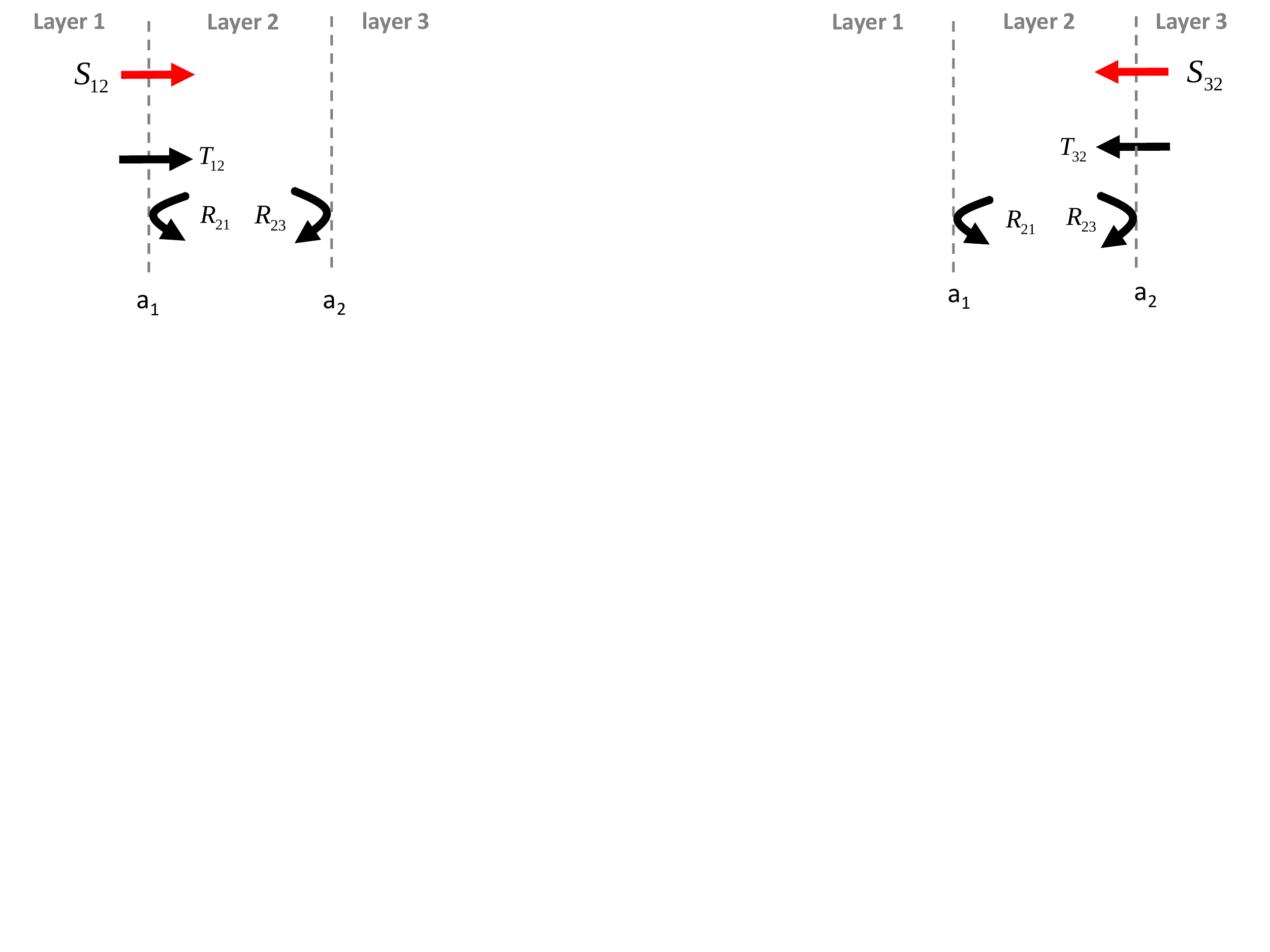}
    }
    \caption{Schematic description of three layers with associated coefficients for the standing-potential case: (a) $R_{32}$ and (b) $S_{32}$.}
    \label{F.3layers.standing}
\end{figure}

\begin{figure}[t]
    \centering
    \includegraphics[width=4.0in]{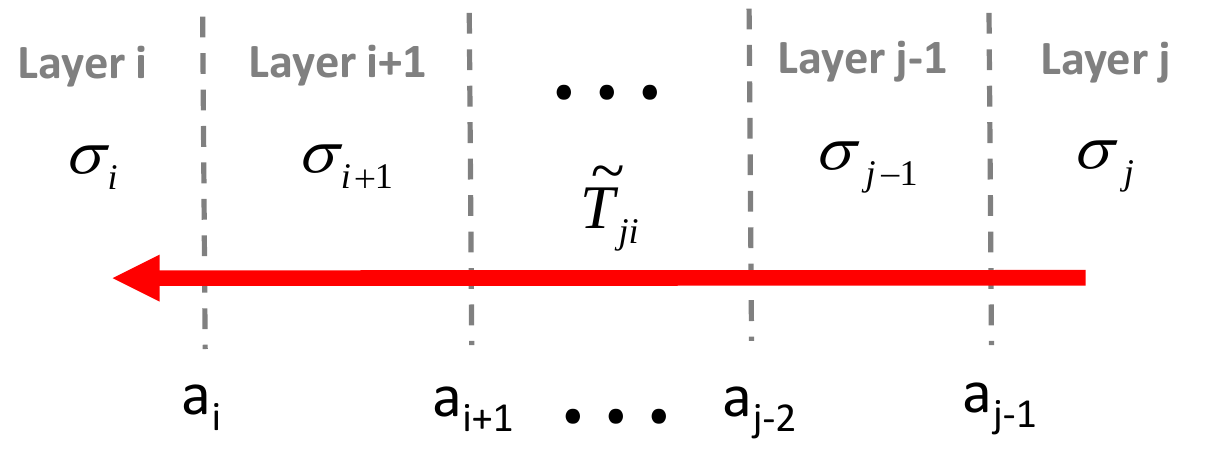}\\
    \caption{Generalized transmission coefficient for the standing-potential case.}
    \label{F.Tjicase2}
\end{figure}

For the standing-potential case depicted in Fig. \ref{F.3layers.standing}, the potentials in three cylindrical layers can be written as
\begin{subequations}
\begin{flalign}
\psi_1&=\In(\lam\rho)B_1, \label{3layers.stand.psi1} \\
\psi_2&=\left[R_{21}\Kn(\lam\rho)+\In(\lam\rho)\right]B_2, \label{3layers.stand.psi2} \\
\psi_3&=\left[\tR_{32}\Kn(\lam\rho)+\In(\lam\rho)\right]B_3. \label{3layers.stand.psi3}
\end{flalign}
\end{subequations}
Similarly, applying proper constraint conditions, we obtain
\begin{flalign}
\tR_{32}=R_{32}+T_{23}R_{21}(1-R_{23}R_{21})^{-1}T_{32}. \label{3layers.stand.R32}
\end{flalign}
and, more generally,
\begin{flalign}
\tR_{i,i-1}=R_{i,i-1}+T_{i-1,i}\tR_{i-1,i-2}(1-R_{i-1,i}\tR_{i-1,i-2})^{-1}T_{i,i-1}.
	\label{multi.layers.stand.Rji}
\end{flalign}
To obtain all amplitudes $B_i$'s, it is convenient to define the $S$ coefficient with decreasing subscripts $S_{i,i-1}$. Since the relation between two successive amplitudes is
\begin{flalign}
B_{i-1}=T_{i,i-1}B_i + R_{i-1,i}R_{i-1,i-2} B_{i-1}, \label{multi.layers.stand.A}
\end{flalign}
the $S$ coefficient in this case is defined as
\begin{flalign}
S_{i,i-1}=(1-R_{i-1,i}\tR_{i-1,i-2})^{-1}T_{i,i-1} \label{multi.layers.stand.S}
\end{flalign}
such that $B_{i-1}=S_{i,i-1}B_i$.
The generalized transmission coefficient depicted in Fig. \ref{F.Tjicase2} can then be written as
\begin{flalign}
\tT_{ji}=T_{i+1,i}S_{i+2,i+1}\cdots S_{j,j-1}
	    =T_{i+1,i}\prod_{k=j}^{i+2}S_{k,k+1}
        =T_{i+1,i}X_{j,i+1}. \label{stand.Tji}
\end{flalign}
Note that $i<j$ for the outgoing-potential case. When $j=i+1$, $X_{j,i+1}=1$ in \eqref{stand.Tji}.

Using generalized reflection and transmission coefficients, we can extend \eqref{psi.off.origin} to incorporate multiple reflections and transmissions. The integral part of \eqref{psi.off.origin} can be modified to
\begin{flalign}
\intzi\left[\In(\lam\rho_<)\Kn(\lam\rho_>)+\Kn(\lam\rho)A_{jn}(\rho')+\In(\lam\rho)B_{jn}(\rho')\right]
	\cos(\lam (z-z'))d\lam, \label{int.part}
\end{flalign}
where the two unknowns are $A_{jn}(\rho')$ and $B_{jn}(\rho')$.
Applying two constraint conditions at $\rho=a_{j-1}$ and $\rho=a_j$, we obtain
\begin{subequations}
\begin{flalign}
A_{jn}(\rho')=(1-\tR_{j,j-1}\tR_{j,j+1})^{-1}\tR_{j,j-1}
	\left[\Kn(\lam\rho')+\In(\lam\rho')\tR_{j,j+1}\right], \label{Ajn} \\
B_{jn}(\rho')=(1-\tR_{j,j+1}\tR_{j,j-1})^{-1}\tR_{j,j+1}
	\left[\In(\lam\rho')+\Kn(\lam\rho')\tR_{j,j-1}\right]. \label{Bjn}
\end{flalign}
\end{subequations}
Substituting \eqref{Ajn} and \eqref{Bjn} into \eqref{int.part} and rearranging the integrand excluding the cosine factor gives
\begin{flalign}
&\left[\In(\lam\rho_<)\Kn(\lam\rho_>)+\Kn(\lam\rho)A_{jn}(\rho')+\In(\lam\rho)B_{jn}(\rho')\right] \notag\\
&=\left\{
\renewcommand{\arraystretch}{2.0}
	\begin{array}{l l}
		\left[\Kn(\lam\rho)+\In(\lam\rho)\tR_{j,j+1}\right]
			\left[\In(\lam\rho')+\Kn(\lam\rho')\tR_{j,j-1}\right]M_j, & \quad \rho>\rho'\\
		\left[\In(\lam\rho)+\Kn(\lam\rho)\tR_{j,j-1}\right]
			\left[\Kn(\lam\rho')+\In(\lam\rho')\tR_{j,j+1}\right]M_j, & \quad \rho<\rho'
	\end{array} \right. \label{int.part.m}
\end{flalign}
where $M_j=\left(1-\tR_{j,j-1}\tR_{j,j+1}\right)^{-1}$. When $i>j$, the potential in layer $i$ is expressed as
\begin{flalign}
\psi_i&=\frac{I}{2\pi^2\sigma_j}\suma e^{\iu n(\phi-\phi')}\intzi
	\left[\Kn(\lam\rho)+\In(\lam\rho)\tR_{i,i+1}\right]A_{in}\cos(\lam (z-z'))d\lam, \label{psi.i.case1}
\end{flalign}
where $A_{in}$ is the amplitude of the outgoing-potential in layer $i$ and expressed as
\begin{flalign}
A_{in}=\left(1-R_{i,i-1}\tR_{i,i+1}\right)\tT_{ji}A_{jn}
	=N_{i+} \tT_{ji}A_{jn}
	=N_{i+} \tT_{ji}\left[\In(\lam\rho')+\Kn(\lam\rho')\tR_{j,j-1}\right]M_j. \label{Ain.case1}
\end{flalign}
Therefore, the integrand factor for layer $i$ in \eqref{psi.i.case1} is
\begin{flalign}
\left[\Kn(\lam\rho)+\In(\lam\rho)\tR_{i,i+1}\right]
	\left[\In(\lam\rho')+\Kn(\lam\rho')\tR_{j,j-1}\right]N_{i+}\tT_{ji}M_j.
 \label{int.part.layer.i.case1}
\end{flalign}
When $i<j$, the potential in layer $i$ is expressed as
\begin{flalign}
\psi_i&=\frac{I}{2\pi^2\sigma_j}\suma e^{\iu n(\phi-\phi')}\intzi
	\left[\In(\lam\rho)+\Kn(\lam\rho)\tR_{i,i-1}\right]A_{in}\cos(\lam (z-z'))d\lam, \label{psi.i.case2}
\end{flalign}
where $A_{in}$ is the amplitude of the standing-potential in layer $i$ and expressed as
\begin{flalign}
A_{in}=\left(1-R_{i,i+1}\tR_{i,i-1}\right)\tT_{ji}A_{jn}
	  =N_{i-} \tT_{ji}A_{jn}\left[\Kn(\lam\rho')+\In(\lam\rho')\tR_{j,j+1}\right]M_j \label{Ain.case2}
\end{flalign}
The integrand of the integral for potential in layer $i$ in \eqref{psi.i.case2} is
\begin{flalign}
\left[\In(\lam\rho)+\Kn(\lam\rho)\tR_{i,i-1}\right]
	\left[\Kn(\lam\rho')+\In(\lam\rho')\tR_{j,j+1}\right]N_{i-}\tT_{ji}M_j.
 \label{int.part.layer.i.case2}
\end{flalign}

In summary, the potential in cylindrically stratified media admits four different expressions depending on the relative position of $\rho$ and $\rho'$, which can be written as
\begin{flalign}
\psi_i&=\frac{\mathcal{I}}{2\pi^2\sigma_j}\suma e^{\iu n(\phi-\phi')}\intzi
	F_n(\rho,\rho')\cos(\lam (z-z'))d\lam, \label{psi.final}
\end{flalign}
where \\
\begin{subequations}
\text{{\it Case 1}: $\rho$ and $\rho'$ are in the same layer and $\rho\geq\rho'$}
\begin{flalign}
&\quad F_n(\rho,\rho')=
	\left[\Kn(\lam\rho)+\In(\lam\rho)\tR_{j,j+1}\right]	
	\left[\In(\lam\rho')+\Kn(\lam\rho')\tR_{j,j-1}\right]M_j,&
    \label{psi.final.case1}
\end{flalign}
\text{{\it Case 2}: $\rho$ and $\rho'$ are in the same layer and $\rho<\rho'$}
\begin{flalign}
&\quad F_n(\rho,\rho')=
	\left[\In(\lam\rho)+\Kn(\lam\rho)\tR_{j,j-1}\right]
	\left[\Kn(\lam\rho')+\In(\lam\rho')\tR_{j,j+1}\right]M_j,&
    \label{psi.final.case2}
\end{flalign}
\text{{\it Case 3}: $\rho$ and $\rho'$ are in different layers and $\rho>\rho'$}
\begin{flalign}
&\quad F_n(\rho,\rho')=
	\left[\Kn(\lam\rho)+\In(\lam\rho)\tR_{i,i+1}\right]
	\left[\In(\lam\rho')+\Kn(\lam\rho')\tR_{j,j-1}\right]N_{i+}\tT_{ji}M_j,&
    \label{psi.final.case3}
\end{flalign}
\text{{\it Case 4}: $\rho$ and $\rho'$ are in different layers and $\rho<\rho'$}
\begin{flalign}
&\quad F_n(\rho,\rho')=
	\left[\In(\lam\rho)+\Kn(\lam\rho)\tR_{i,i-1}\right]
	\left[\Kn(\lam\rho')+\In(\lam\rho')\tR_{j,j+1}\right]N_{i-}\tT_{ji}M_j.&
    \label{psi.final.case4}
\end{flalign}
\end{subequations}

\subsection{Rescaled modified-Bessel functions}
\label{sec.2.3}
The electric potential involves products of the modified-Bessel function of the first and second kind, viz. $\In$ and $\Kn$. Those products can involve disparate values due to the exponential behavior of the functions. For example, when $|z|<<1$, $\Kn(z)$ has very large value whereas $\In(z)$ has very small value. This disparity becomes progressively worse for higher order modes. On the other hand, when $\Re e[z]>>1$, $\Kn(z)$ has very small value while $\In(z)$ has very large value. This leads to unreliable results under double-precision computations. To eliminate this problem, a new set of rescaled modified-Bessel functions are defined in a similar fashion to what has been done in \cite{Moon14:Stable} for Bessel and Hankel functions, viz. $\Jn$ and $\Hn$. In order to apply such rescaled functions, the analytical expressions for the potential need to modified accordingly, as described next.

When $|z| \ll 1$, $\In(z)$ and $\Kn(z)$ can be expressed via small argument approximations for $n>0$.
Noting that, for $-\pi<arg(z)<\pi/2$, the relationship between cylindrical functions and modified-cylindrical functions reads as
\begin{subequations}
\begin{flalign}
\In(z)=\iu^{-n}\Jn(iz), \label{In.Jn}\\
\Kn(z)=\frac{\pi}{2}\iu^{n+1}\Hn(iz). \label{Kn.Hn}
\end{flalign}
\end{subequations}
Thus, we have the following small argument approximations for $\In$ and $\Kn$ and their derivatives.
\begin{subequations}
\begin{flalign}
\In(\lam a_i)&\approx\frac{1}{n!}\left(\frac{\lam a_i}{2}\right)^n=
    \frac{1}{n!}\left(\frac{\lam}{2}\right)^n\cdot a_i^n\cdot 1=
    G a_i^n\hIn(\lam a_i), \label{In.small}\\
\Ind(\lam a_i)&\approx\frac{1}{2(n-1)!}\left(\frac{\lam a_i}{2}\right)^{n-1}=
    \frac{1}{n!}\left(\frac{\lam}{2}\right)^n\cdot a_i^n\cdot \frac{n}{\lam a_i}=
    G a_i^n\hInd(\lam a_i), \label{Ind.small}\\
\Kn(\lam a_i)&\approx \frac{(n-1)!}{2}\left(\frac{2}{\lam a_i}\right)^n=
    n!\left(\frac{2}{\lam}\right)^n\cdot a_i^{-n}\cdot \left(\frac{1}{2n}\right)=
    G^{-1}a_i^{-n}\hKn(\lam a_i), \label{Kn.small}\\
\Knd(\lam a_i)&\approx-\frac{n!}{4}\left(\frac{2}{\lam a_i}\right)^{n+1}=
    n!\left(\frac{2}{\lam}\right)^n\cdot a_i^{-n}\cdot \left(-\frac{1}{2\lam a_i}\right)=
    G^{-1}a_i^{-n}\hKnd(\lam a_i). \label{Knd.small}
\end{flalign}
\end{subequations}
It should be noted that the multiplicative factor $G$ above is chosen so as to not depend on the radial distance, $a_i$. Also, $G$ is identical for a function and its derivative, and the multiplicative factors appearing in $\In$ and $\Kn$ are reciprocal to each other. This will facilitate some computations later on.

When $\Re e[z] \gg 1$ and $z=\lam a_i=(\lam'+\iu\lam'')a_i$, the large argument approximations for the modified-Bessel functions write as \cite{Jin:SpecialFunctions}
\begin{subequations}
\begin{flalign}
\In(\lam a_i)&=\frac{e^{\lam a_i}}{\sqrt{2\pi\lam a_i}}\left[1-\frac{(\mu-1)}{1!(8\lam a_i)}
    +\frac{(\mu-1)(\mu-9)}{2!(8\lam a_i)^2}+\dotsb \right] \notag\\
     &=e^{\lam' a_i}\hIn(\lam a_i), \label{In.large}\\
\Kn(\lam a_i)&=\sqrt{\frac{\pi}{2\lam a_i}}e^{-\lam a_i}\left[1+\frac{(\mu-1)}{1!(8\lam a_i)}
    +\frac{(\mu-1)(\mu-9)}{2!(8\lam a_i)^2}+\dotsb \right] \notag\\
     &=e^{-\lam' a_i}\hKn(\lam a_i), \label{Kn.large}
\end{flalign}
\end{subequations}
where $\mu=4n^2$. Again, the associated multiplicative factors are reciprocal to each other. The derivatives of rescaled modified-cylindrical functions for large arguments can be derived through the recursive formulas such that
\begin{flalign}
\Ind(\lam a_i)
	&=I_{n-1}(\lam a_i)-\frac{n}{\lam a_i} \In(\lam a_i)
	=e^{\lam' a_i}\hat{I}_{n-1}(\lam a_i) - e^{\lam' a_i}\frac{n}{\lam a_i}\hIn(\lam a_i)
	=e^{\lam' a_i}\hInd(\lam a_i), \label{Ind.large}\\
\Knd(\lam a_i)
	&=-K_{n-1}(\lam a_i)-\frac{n}{\lam a_i}\Kn(\lam a_i)
	=-e^{-\lam' a_i}\hat{K}_{n-1}(\lam a_i) - e^{-\lam' a_i}\frac{n}{\lam a_i}\hKn(\lam a_i)
	=e^{-\lam' a_i}\hKnd(\lam a_i). \label{Knd.large}
\end{flalign}

If the argument is neither small nor large, rescaled modified-cylindrical functions are defined, in analogy to small and large arguments, as
\begin{subequations}
\begin{flalign}
\In(\lam a_i)&=P_i\hIn(\lam a_i), \label{In.mod}\\
\Ind(\lam a_i)&=P_i\hInd(\lam a_i), \label{Ind.mod}\\
\Kn(\lam a_i)&=P_i^{-1}\hKn(\lam a_i), \label{Kn.mod}\\
\Knd(\lam a_i)&=P_i^{-1}\hKnd(\lam a_i), \label{Knd.mod}
\end{flalign}
\end{subequations}
where the multiplicative factor $P_i$ is defined as in~\cite{Moon14:Stable}, i.e.,
\begin{subequations}
\begin{flalign}
&\text{If } |\In(\lam a_i)|^{-1} < T_m, \quad P_i = 1. \label{Pi.a}\\
&\text{If } |\In(\lam a_i)|^{-1} \geq T_m,
	\quad P_i = |\In(\lam a_i)|. \label{Pi.b}
\end{flalign}
\end{subequations}
with $T_m = 10^100$ used for double-precision arithmetic computations.

and its subscript is linked to $a_i$.
As Table \ref{T.RCCMF} shows, the argument for rescaled modified-cylindrical functions can be categorized into small, moderate, and large, with different and appropriate multiplicative factors defined accordingly.
\begin{table}[t]
\begin{center}
\renewcommand{\arraystretch}{1.8}
\setlength{\tabcolsep}{8pt}
\caption{Definition of rescaled modified-cylindrical functions for all types of arguments.}
\begin{tabular}{|c|c|c|c|}
    \hline
     & small arguments & moderate arguments & large arguments \\
    \hline\hline
    $\In(\lam a_i)$ & $G a_i^n\hIn(\lam a_i)$ &
    $P_i\hIn(\lam a_i)$ & $e^{\lam' a_i}\hIn(\lam a_i)$ \\
     \hline
    $\Ind(\lam a_i)$ & $G a_i^n\hInd(\lam a_i)$ &
    $P_i\hInd(\lam a_i)$ & $e^{\lam' a_i}\hInd(\lam a_i)$ \\
     \hline
    $\Kn(\lam a_i)$ & $G^{-1} a_i^{-n}\hKn(\lam a_i)$ &
    $P_i^{-1}\hKn(\lam a_i)$ & $e^{-\lam' a_i}\hKn(\lam a_i)$ \\
     \hline
    $\Knd(\lam a_i)$ & $G^{-1} a_i^{-n}\hKnd(\lam a_i)$ &
    $P_i^{-1}\hKnd(\lam a_i)$ & $e^{-\lam' a_i}\hKnd(\lam a_i)$ \\
    \hline
\end{tabular}
\label{T.RCCMF}
\end{center}
\end{table}

The numerical threshold values used here to define small, moderate, and large argument ranges are identical to those used for the time-harmonic case involving cylindrical Bessel and Hankel functions detailed in \cite{Moon14:Stable} and not repeated here\footnote{A third type of threshold, used in connection with modified-Bessel functions {\it magnitudes} (not arguments) is also necessary within the moderate argument range. Again, this threshold value is identical to the one applied to ordinary Bessel functions in \cite{Moon14:Stable}}.

\subsection{Rescaled reflection and transmission coefficients}
\label{sec.2.5}

We can classify the multiplicative factors associated with the rescaled modified-cylindrical functions into two types, denoted as $\alpha$ and $\beta$, and shown in Table \ref{T.alpha.beta}. The factor $\alpha_i$ is associated with $\Kn(\lam a_i)$, whereas $\beta_i$ is associated with $\In(\lam a_i)$. Note again that the subscript $i$ refers to to the index of the radial factor $a_i$ in the argument. There are two important properties to note for $\alpha$ and $\beta$:
($i$){\it Reciprocity}:
$\alpha_i=1/\beta_i$ and
($ii$) {\it Boundness}: $
|\beta_i\alpha_j|\leq 1, \quad\text{for } i < j$.
As we will see below, these two properties are important in ensuring stable numerical computations.

\begin{table}[h]
\begin{center}
\renewcommand{\arraystretch}{1.3}
\setlength{\tabcolsep}{8pt}
\caption{Definition of $\alpha_i$ and $\beta_i$.}
\begin{tabular}{|c|c|c|}
    \hline
    argument type & $\alpha_i$ & $\beta_i$ \\
    \hline
    small  & $G^{-1} a^{-n}_i$ & $G a^n_i$ \\
    moderate & $P_i^{-1}$ & $P_i$\\
    large  & $e^{-\lam' a_i}$ & $e^{\lam' a_i}$\\
    \hline
\end{tabular}
\label{T.alpha.beta}
\end{center}
\end{table}

Recalling the expressions for the reflection and transmission coefficients obtained before, the reflection coefficient $R_{12}$ for the outgoing-potential case is modified to
\begin{flalign}
R_{12}&=\frac{(\sigma_2-\sigma_1)\Kn(\lam a_1)\Knd(\lam a_1)}
	{\sigma_1\Ind(\lam a_1)\Kn(\lam a_1) - \sigma_2\In(\lam a_1)\Knd(\lam a_1)} \notag\\
      &=\alpha_1^2\frac{(\sigma_2-\sigma_1)\hKn(\lam a_1)\hKnd(\lam a_1)}
	{\sigma_1\hInd(\lam a_1)\hKn(\lam a_1) - \sigma_2\hIn(\lam a_1)\hKnd(\lam a_1)} \notag\\
	  &=\alpha_1^2\hR_{12}.  \label{R12.cond}
\end{flalign}
Similarly, it can be shown that the reflection coefficient $R_{21}$ for the standing-potential case is modified to
$
R_{21}=\beta_1^2\hR_{21}$, and that
the transmission coefficient $T_{12}$ for the outgoing-potential case
and the transmission coefficient $T_{21}$ for the standing-potential case
simply recover the original ones without any multiplicative factors, i.e. $T_{12}=\hT_{12}$
and $T_{21}=\hT_{21}$.

Based on the above modifications for the reflection and transmission coefficients, generalized reflection and transmission coefficients for thre or more layers can be similarly modified. After some algebra, it can be shown that
$\tR_{i,i+1}=\alpha_i^2\htR_{i,i+1}$ and
$\tR_{i+1,i}=\beta_i^2\htR_{i+1,i}$, and that, for both the outgoing-potential and standing-potential cases,
$\tT_{ij}=\htT_{ij}$.

In addition to generalized reflection and transmission coefficients, the factors $M_j$ and $N_{i\pm}$ considered before are also required to compute the potential. The basic difference between the two types of coefficients is that $M_j$ involves two generalized reflection coefficients whereas $N_{i\pm}$ involves only one generalized reflection coefficient. All these auxiliary coefficients can be redefined accordingly using rescaled reflection coefficients, i.e.,
\begin{subequations}
\begin{flalign}
M_j&=
    \left[1-\tR_{j,j-1}\tR_{j,j+1}\right]^{-1}=
    \left[1-\beta^2_{j-1}\alpha^2_j\htR_{j,j-1}\htR_{j,j+1}\right]^{-1},
    \label{Mj}\\
N_{i+}&=
    \left[1-R_{i,i-1}\tR_{i,i+1}\right]^{-1}=
    \left[1-\beta^2_{i-1}\alpha^2_i\hR_{i,i-1}\htR_{i,i+1}\right]^{-1},
    \label{Ni.plus}\\
N_{i-}&=
    \left[1-R_{i,i+1}\tR_{i,i-1}\right]^{-1}=
    \left[1-\beta^2_{i-1}\alpha^2_i\hR_{i,i+1}\htR_{i,i-1}\right]^{-1}.
     \label{Ni.minus}
\end{flalign}
\end{subequations}

\subsection{Rescaled integrand}
\label{sec.2.6}
Nest step is to modify the full integrand using rescaled modified-cylindrical functions. Since there are four integrand expressions, depending of the relative position of $\rho$ and $\rho'$, each case is considered separately.

For {\it Case 1}, there are four radial parameters of interest: $a_{j-1}$, $\rho'$, $\rho$, and $a_j$. For convenience, we let $a_{j-1}=a_1$, $\rho'=a_2$, $\rho=a_3$, and $a_j=a_4$ so that $a_1<a_2<a_3<a_4$, and the integrand rewrites as
\begin{flalign}
F_n(\rho,\rho')
&=\left[\Kn(\lam\rho) + \In(\lam\rho)\tR_{j,j+1}\right]	
  \left[\In(\lam\rho') + \Kn(\lam\rho')\tR_{j,j-1}\right]M_j \notag\\
&=\left[\beta_2\alpha_3\hKn(\lam\rho) + (\beta_2\alpha_4)(\beta_3\alpha_4)\hIn(\lam\rho)\htR_{j,j+1}\right]	
  \left[\hIn(\lam\rho') + (\beta_1\alpha_2)^2\hKn(\lam\rho')\htR_{j,j-1}\right]M_j \notag\\
&=\left[A_1\hKn(\lam\rho) + A_2\hIn(\lam\rho)\htR_{j,j+1}\right]	
  \left[A_3\hIn(\lam\rho') + A_4\hKn(\lam\rho')\htR_{j,j-1}\right]M_j, \label{Fn.case1.cond}
\end{flalign}
where $\beta_2^{-1}=\alpha_2$ has been used. All multiplicative factors $A_1$, $A_2$, $A_3$, $A_4$ have magnitudes no larger than one due to the boundness property discussed above.

 Similarly, for {\it Case 2}, there are four radial parameters of interest: $a_{j-1}$, $\rho$, $\rho'$, and $a_j$. For convenience, we let $a_{j-1}=a_1$, $\rho=a_2$, $\rho'=a_3$, and $a_j=a_4$ so that $a_1<a_2<a_3<a_4$, and the integrand rewrites as
\begin{flalign}
F_n(\rho,\rho')
&=\left[\In(\lam\rho) + \Kn(\lam\rho)\tR_{j,j-1}\right]
  \left[\Kn(\lam\rho') + \In(\lam\rho')\tR_{j,j+1}\right]M_j \notag\\
&=\left[\beta_2\alpha_3\hIn(\lam\rho) + (\beta_1\alpha_2)(\beta_1\alpha_3)\hKn(\lam\rho)\htR_{j,j-1}\right]
  \left[\hKn(\lam\rho') + (\beta_3\alpha_4)^2\hIn(\lam\rho')\htR_{j,j+1}\right]M_j \notag\\
&=\left[B_1\hIn(\lam\rho) + B_2\hKn(\lam\rho)\htR_{j,j-1}\right]
  \left[B_3\hKn(\lam\rho') + B_4\hIn(\lam\rho')\htR_{j,j+1}\right]M_j. \label{Fn.case2.cond}
\end{flalign}
Again, all multiplicative factors $B_1$, $B_2$, $B_3$, $B_4$ magnitudes are bounded by one.

For {\it Case 3}, there are six radial parameters of interest: $a_{j-1}$, $\rho'$, $a_j$, $a_{i-1}$, $\rho$, and $a_i$. For convenience, we let $a_{j-1}=a_1$, $\rho'=a_2$, $a_j=a_3$, $a_{i-1}=a_4$, $\rho=a_5$, and $a_i=a_6$ so that so that $a_1<a_2<a_3\le a_4<a_5<a_6$, and the integrand rewrites as
\begin{flalign}
F_n(\rho,\rho')
&=\left[\Kn(\lam\rho) + \In(\lam\rho)\tR_{i,i+1}\right]
  \left[\In(\lam\rho') + \Kn(\lam\rho')\tR_{j,j-1}\right]N_{i+}\tT_{ji}M_j \notag\\
&=\left[\beta_2\alpha_5\hKn(\lam\rho) + (\beta_2\alpha_6)(\beta_5\alpha_6)\hIn(\lam\rho)\htR_{i,i+1}\right]
  \left[\hIn(\lam\rho') + (\beta_1\alpha_2)^2\hKn(\lam\rho')\htR_{j,j-1}\right]N_{i+}\tT_{ji}M_j \notag\\
&=\left[C_1\hKn(\lam\rho) + C_2\hIn(\lam\rho)\htR_{i,i+1}\right]
  \left[C_3\hIn(\lam\rho') + C_4\hKn(\lam\rho')\htR_{j,j-1}\right]N_{i+}\tT_{ji}M_j. \label{Fn.case3.cond}
\end{flalign}
All multiplicative factors $C_1$, $C_2$, $C_3$, $C_4$ have magnitudes never greater than unity.

Finally, for {\it Case 4}, there are again six radial parameters of interest: $a_{i-1}$, $\rho$, $a_i$, $a_{j-1}$, $\rho'$, and $a_j$. For convenience, we let $a_{i-1}=a_1$, $\rho=a_2$, $a_i=a_3$, $a_{j-1}=a_4$, $\rho'=a_5$, and $a_j=a_6$ so that $a_1<a_2<a_3\le a_4<a_5<a_6$, and the integrand rewrites as
\begin{flalign}
F_n(\rho,\rho')
&=\left[\In(\lam\rho) + \Kn(\lam\rho)\tR_{i,i-1}\right]
  \left[\Kn(\lam\rho') + \In(\lam\rho')\tR_{j,j+1}\right]N_{i-}\tT_{ji}M_j \notag\\
&=\left[\beta_2\alpha_5\hIn(\lam\rho) + (\beta_1\alpha_2)(\beta_1\alpha_5)\hKn(\lam\rho)\htR_{i,i-1}\right]
  \left[\hKn(\lam\rho') + (\beta_5\alpha_6)^2\hIn(\lam\rho')\htR_{j,j+1}\right]N_{i-}\tT_{ji}M_j \notag\\
&=\left[D_1\hIn(\lam\rho) + D_2\hKn(\lam\rho)\htR_{i,i-1}\right]
  \left[D_3\hKn(\lam\rho') + D_4\hIn(\lam\rho')\htR_{j,j+1}\right]N_{i-}\tT_{ji}M_j. \label{Fn.case4.cond}
\end{flalign}
Once more, all multiplicative factors $D_1$, $D_2$, $D_3$, $D_4$ have magnitudes never greater than one.

\subsection{Numerical integration}
\label{sec.2.7}
The electric potential expression includes a semi-infinite integral and an infinite series summation, as shown in \eqref{psi.final}. Therefore, truncation errors are inevitable and an error analysis should be made to ensure reliable results. In addition, the involved Sommerfeld-type integrals can be notoriously slowly convergent. A variety of extrapolation methods are applied here in order to accelerate the numerical integration, as described in more detail in Appendix B.

\section{Results}
\label{sec.4.results}
\begin{figure}[!htbp]
    \centering
    \includegraphics[width=1.5in]{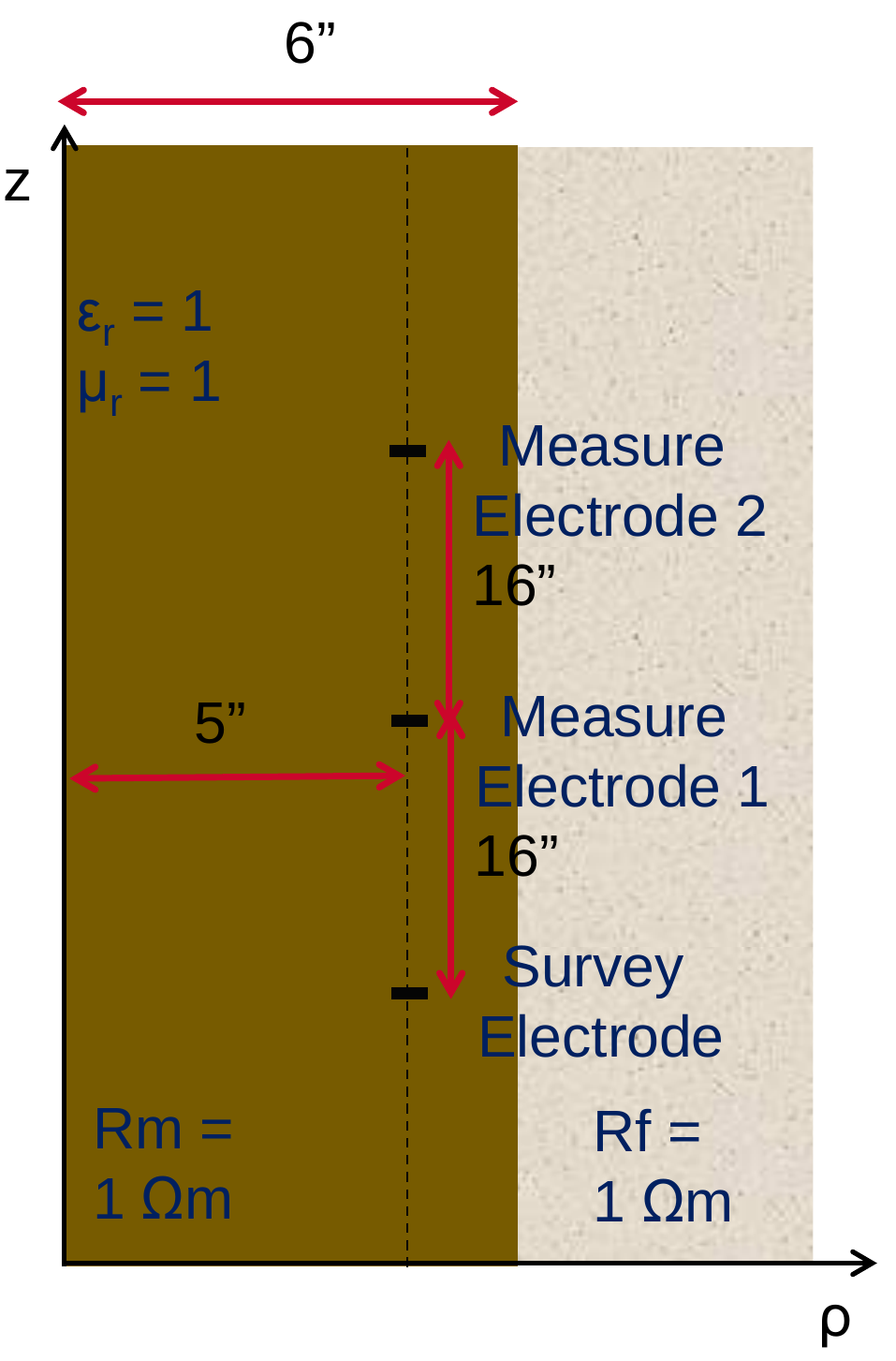}\\
    \caption{Case 1 in the $\rho z$-plane with $\mathcal{R}_1=\mathcal{R}_2=1$.}
    \label{F.case1}
\end{figure}

A number of cases of practical interest for geophysical exploration are considered next. The results below are obtained on
2.6 GHz Opteron with 8 cores and 32 GB memory. For all cases below, the relative permittivity $\epsilon_r$ and relative permeability $\mu_r$ are set to one. The innermost layer has $6^{\prime\prime}$ radius and represents a mud-filled borehole. The outer layers represent the adjacent Earth formations, invasion zones, and/or casing layers.
Each layer assumes different resistivity values $\mathcal{R}$, where $\mathcal{R}=1/\sigma$.
As Fig. \ref{F.case1} illustrates, a current source emitting a DC current of 1 A is located at the Survey Electrode position. The electric potential is measured at two points: the Measurement Electrode 1 ($V_{16^{\prime\prime}}$) positioned 16 inches away from the source along the vertical direction and the Measurement Electrode 2 ($V_{32^{\prime\prime}}$) positioned 32 inches away from the source along the vertical direction. Both the source and two measurement electrodes are spaced 5 inches away from the $z$-axis and their azimuthal positions are the same ($\phi=\phi'=0$). The relevant error parameters as defined in the Appendix and used in the following are $e_{tol}=10^{-4}$ and $e_{thr}=10^{-4}$. The electric potential and resistivity units used here are $[V]$ and $[\Omega\cdot m]$, respectively. The results of the present algorithm are compared against those from the finite element method (FEM).

\subsection{Logging simulation results}
\label{sec.4.1}
Case 1 corresponds to a homogeneous problem, where analytical solutions are available. As Table \ref{T.case1} shows, the new algorithm produces very accurate results with very fast computing time. This is very important for making feasible the solution of the inverse problem (i.e., determining the unknown resistivity profile from known electric potential at the received electrodes) using iterative methods predicated on repeated forward solves.

\vspace{4.0em}
\begin{table}[!htbp]
\begin{center}
\renewcommand{\arraystretch}{1.2}
\setlength{\tabcolsep}{14pt}
\caption{Comparison of electric potential for Case 1.}
    \begin{tabular}{|c|c|c|c|}
        \hline
         & Analytical & FEM & Present Algorithm \\
        \hline
        $V_{16^{\prime\prime}}$ & 1.9581 $\times\;10^{-1}$ & 1.7878 $\times\;10^{-1}$
			& 1.9580 $\times\;10^{-1}$ (2 sec.)\\
        $V_{32^{\prime\prime}}$ & 9.7905 $\times\;10^{-2}$ & 9.4416 $\times\;10^{-2}$
			& 9.7900 $\times\;10^{-2}$ (2 sec.)\\
		\hline
        $\Delta V$              & 9.7905 $\times\;10^{-2}$ & 8.4371 $\times\;10^{-2}$
			& 9.7902 $\times\;10^{-2}$\\
        \hline
    \end{tabular}
	\label{T.case1}
\end{center}
\end{table}

\pagebreak

Cases 2 and 3 correspond to a two-layer problem with 1:5 contrast between the adjacent borehole and formation resistivities.

\begin{figure}[!htbp]
	\centering
	\subfloat[\label{F.case4}]{%
      \includegraphics[width=1.5in]{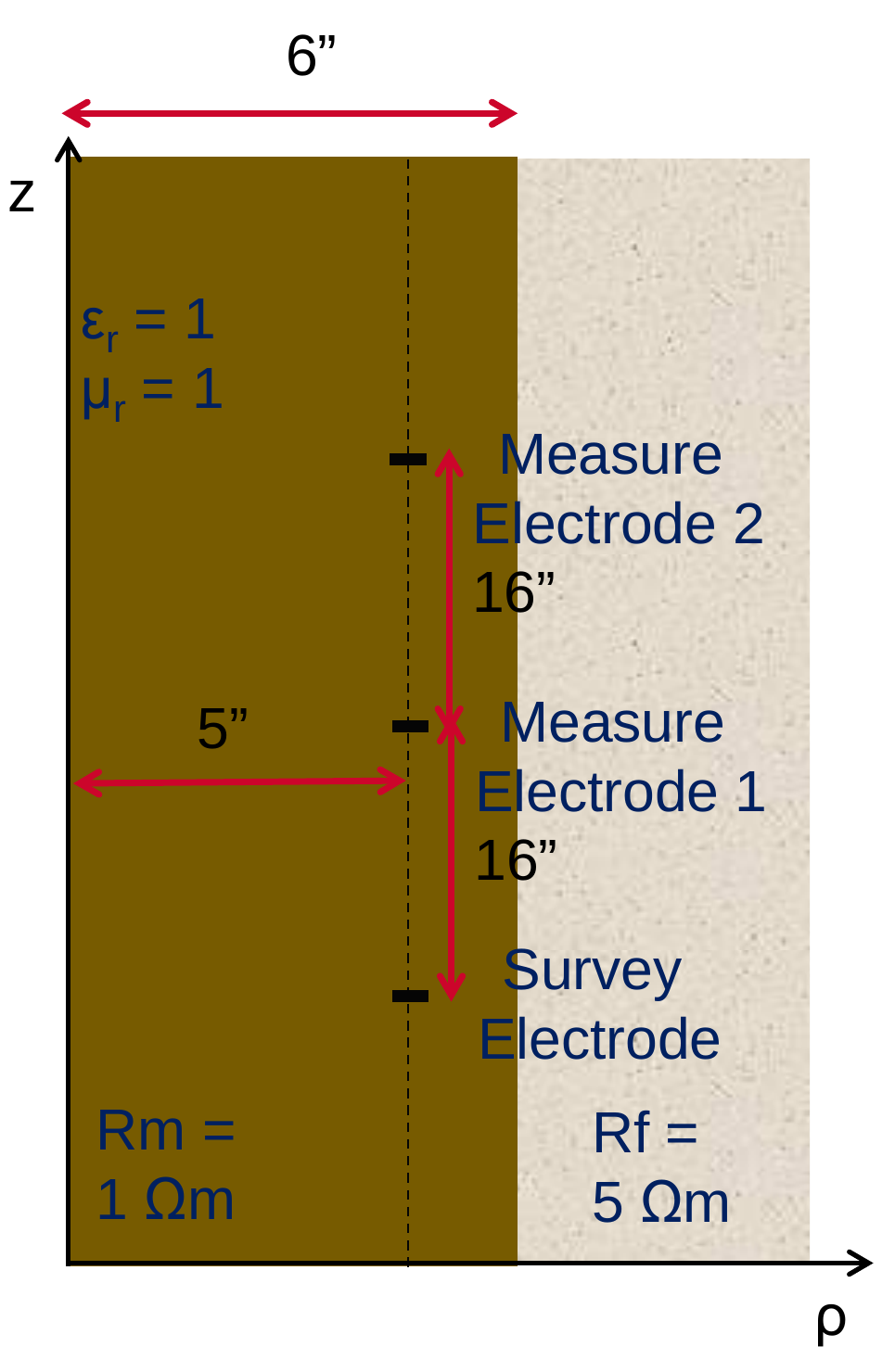}
    }
    \hspace{2.5cm}
    \subfloat[\label{F.case5}]{%
      \includegraphics[width=1.5in]{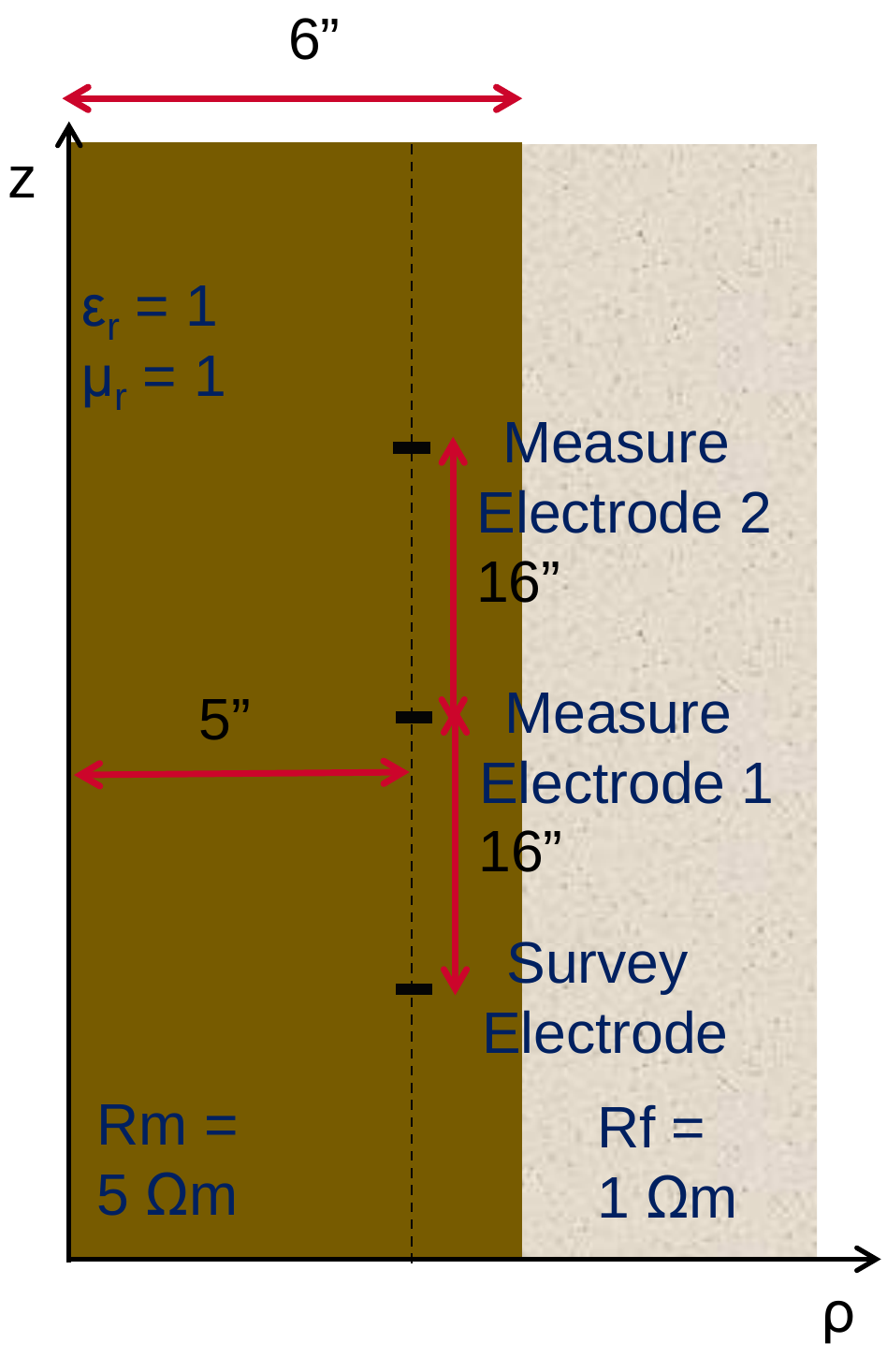}
    }
    \caption{Cases 2 and 3 in the $\rho z$-plane: (a) Case 2 with $\mathcal{R}_1=1$, $\mathcal{R}_2=5$ and (b) Case 3 with $\mathcal{R}_1=5$, $\mathcal{R}_2=1$.}
    \label{F.case45}
\end{figure}
\begin{table}[!htbp]
\begin{center}
\renewcommand{\arraystretch}{1.2}
\setlength{\tabcolsep}{14pt}
\caption{Comparison of electric potential for Case 2.}
    \begin{tabular}{|c|c|c|}
        \hline
         & FEM & Present Algorithm \\
        \hline
        $V_{16^{\prime\prime}}$ & 9.5544 $\times\;10^{-1}$ & 9.7802 $\times\;10^{-1}$  (2 sec.)\\
        $V_{32^{\prime\prime}}$ & 5.4241 $\times\;10^{-1}$ & 5.4981 $\times\;10^{-1}$  (2 sec.)\\
        \hline
        $\Delta V$              & 4.1303 $\times\;10^{-1}$ & 4.2822 $\times\;10^{-1}$\\
        \hline
    \end{tabular}
    \label{T.case4}
\vspace{4em}
\caption{Comparison of electric potential for Case 3.}
    \begin{tabular}{|c|c|c|}
        \hline
         & FEM & Present Algorithm \\
        \hline
        $V_{16^{\prime\prime}}$ & 1.6945 $\times\;10^{-1}$ & 2.0533 $\times\;10^{-1}$  (2 sec.)\\
        $V_{32^{\prime\prime}}$ & 9.0310 $\times\;10^{-2}$ & 9.7677 $\times\;10^{-2}$  (2 sec.)\\
        \hline
        $\Delta V$              & 7.9145 $\times\;10^{-2}$ & 1.0766 $\times\;10^{-1}$\\
        \hline
    \end{tabular}
    \label{T.case5}
\end{center}
\end{table}

\pagebreak

Cases 4 and 5 include a highly conductive casing. As Table \ref{T.case7} shows, there is disagreement between the semi-analytical and FEM results w.r.t. the {\it absolute} value of the electric potentials, although the computed potential {\it differences} (voltage drop) between the two electrodes are very similar. This disagreement is easy to explain as a FEM mesh truncation effect. The electric current in this case flows primarily along the thin, highly conductive casing, which does not produce sufficient decay on the electric potential before before it reaches the mesh boundary. Consequently, the FEM result has a spurious potential offset. The FEM has difficulty in simulating this problem unless a mesh truncation treatment is included on both top and bottom boundaries and/or a very long mesh is used along the $z$ direction. Notwithstanding such discrepancy, note that, in a resistivity logging context, the primary quantity of interest is the potential difference between electrodes.

\begin{figure}[!htbp]
	\centering
	\subfloat[\label{F.case6}]{%
      \includegraphics[width=1.5in]{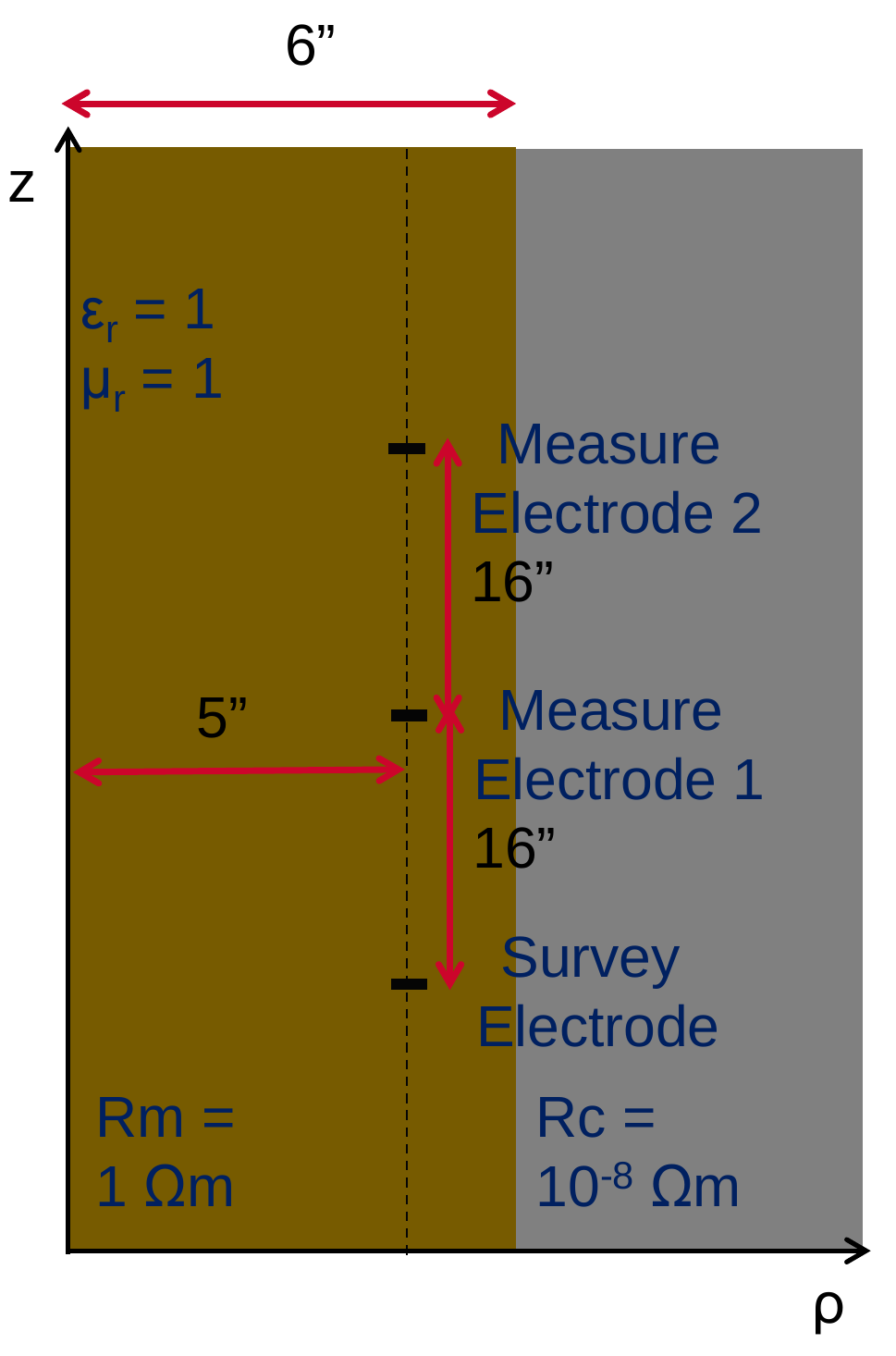}
    }
    \hspace{2.5cm}
    \subfloat[\label{F.case7}]{%
      \includegraphics[width=1.5in]{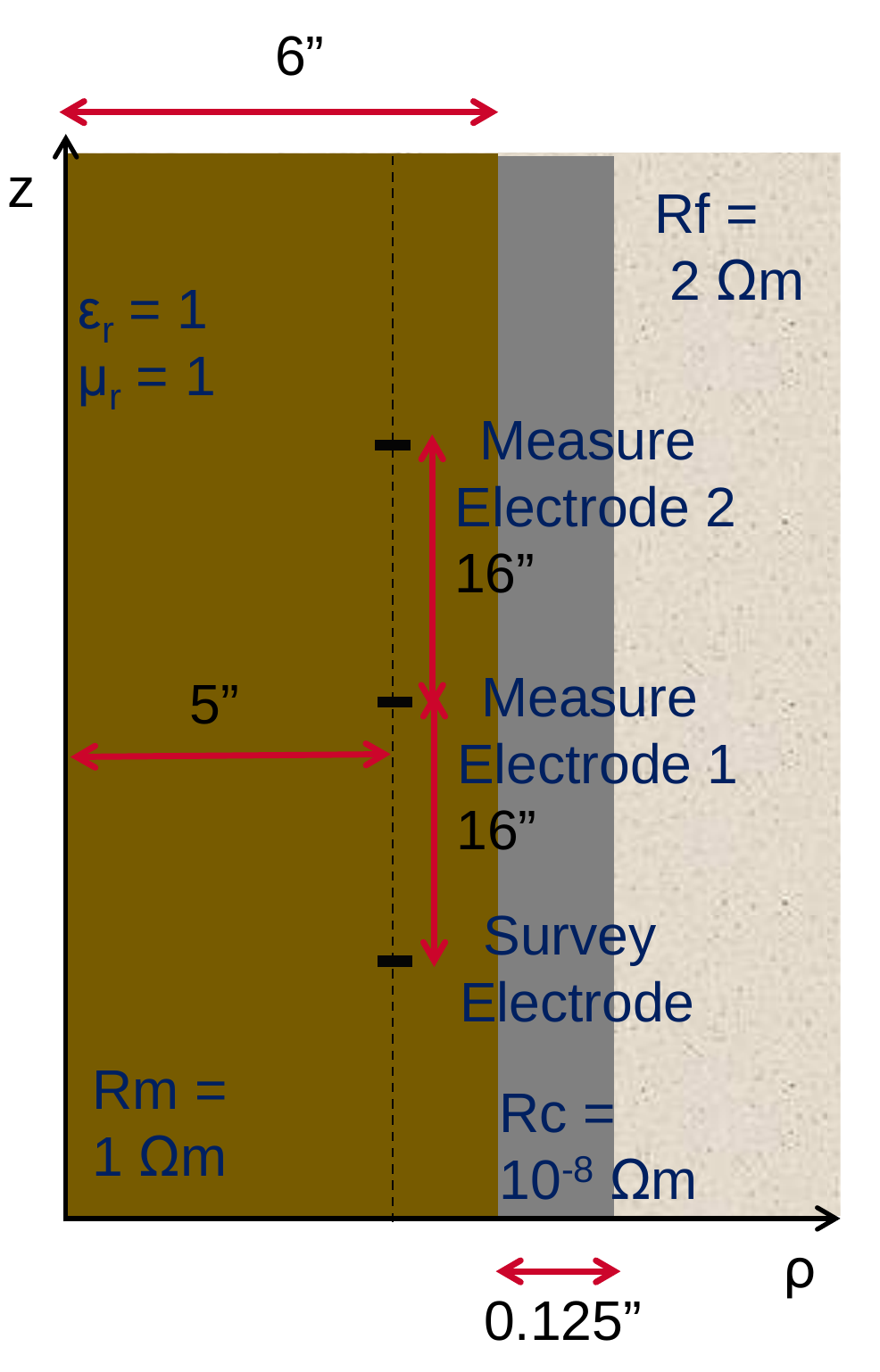}
    }
    \caption{Cases 4 and 5 in the $\rho z$-plane: (a) Case 4 with $\mathcal{R}_1=1$, $\mathcal{R}_2=10^{-8}$ and (b) Case 5 with $\mathcal{R}_1=1$, $\mathcal{R}_2=10^{-8}$, $R_3=2$.}
    \label{F.case67}
\end{figure}
\begin{table}[!htbp]
\begin{center}
\renewcommand{\arraystretch}{1.2}
\setlength{\tabcolsep}{14pt}
\caption{Comparison of electric potential for Case 4.}
    \begin{tabular}{|c|c|c|}
        \hline
         & FEM & Present Algorithm \\
        \hline
        $V_{16^{\prime\prime}}$ & 1.2980 $\times\;10^{-4}$ & 1.3873 $\times\;10^{-4}$ (2 sec.)\\
        $V_{32^{\prime\prime}}$ & 2.1372 $\times\;10^{-7}$ & 2.1415 $\times\;10^{-7}$ (2 sec.)\\
        \hline
        $\Delta V$              & 1.2959 $\times\;10^{-4}$ & 1.3852 $\times\;10^{-4}$\\
        \hline
    \end{tabular}
    \label{T.case6}
\vspace{1em}
\caption{Comparison of electric potential for Case 5.}
    \begin{tabular}{|c|c|c|}
        \hline
         & FEM & Present Algorithm \\
        \hline
        $V_{16^{\prime\prime}}$ & 2.2658 $\times\;10^{-4}$ & 1.7241 $\times\;10^{-3}$ (3 sec.)\\
        $V_{32^{\prime\prime}}$ & 9.6331 $\times\;10^{-5}$ & 1.5885 $\times\;10^{-3}$ (2 sec.)\\
        \hline
        $\Delta V$              & 1.3025 $\times\;10^{-4}$ & 1.3562 $\times\;10^{-4}$\\
        \hline
    \end{tabular}
    \label{T.case7}
\end{center}
\end{table}

\pagebreak

\begin{figure}[!htbp]
	\centering
	\subfloat[\label{F.case8}]{%
      \includegraphics[width=1.5in]{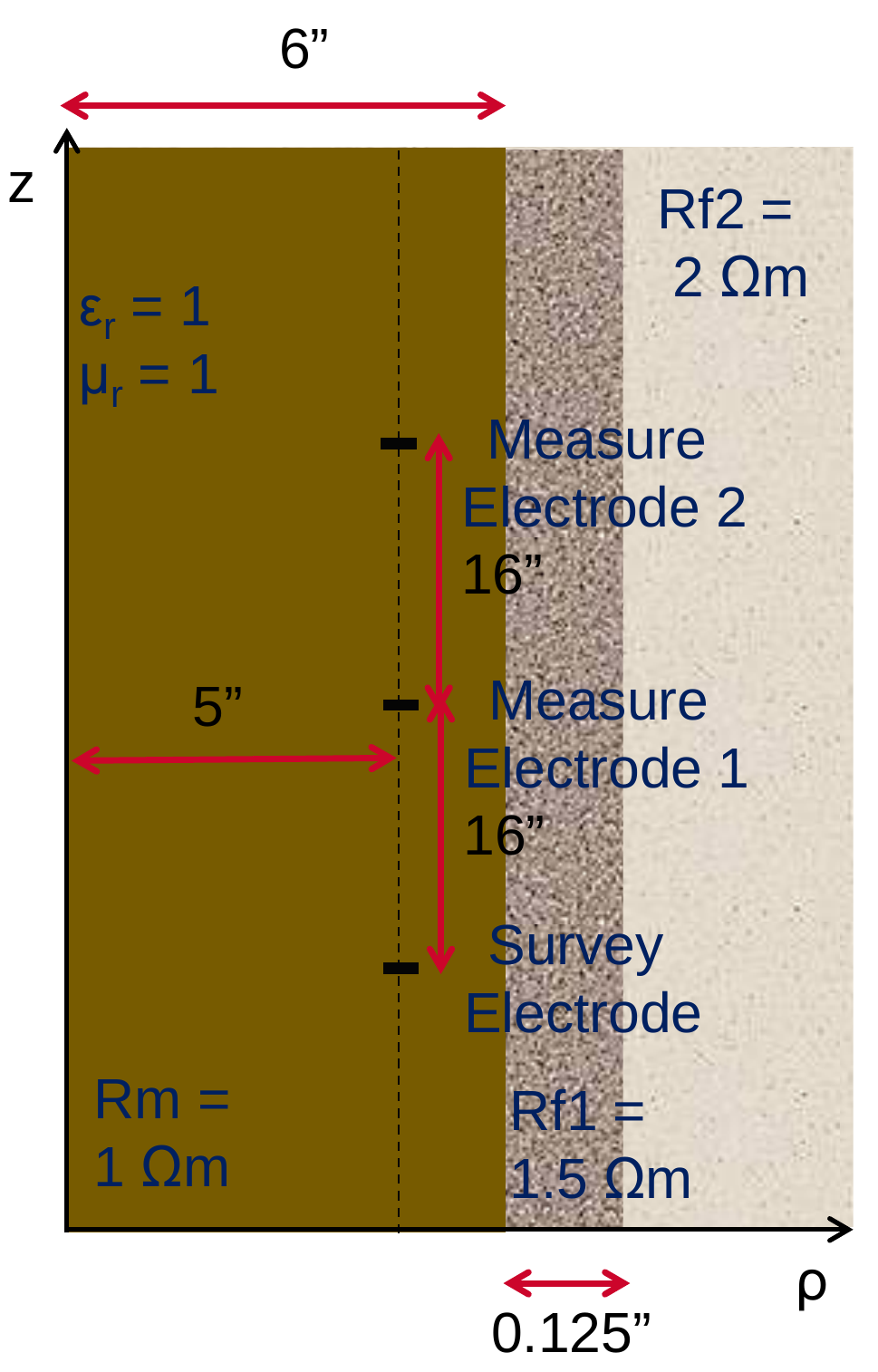}
    }
    \hspace{2.5cm}
    \subfloat[\label{F.case9}]{%
      \includegraphics[width=1.6in]{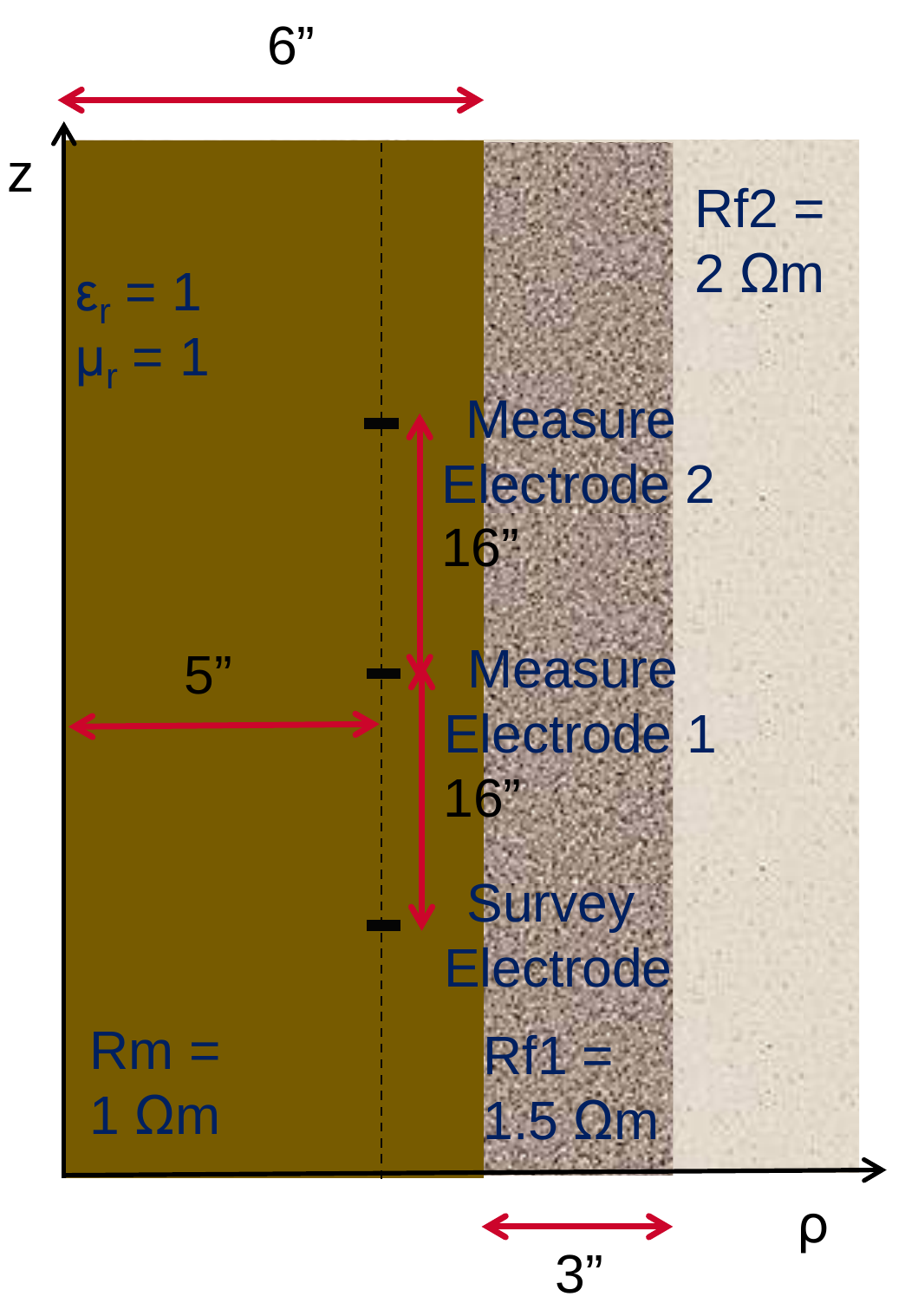}
    }
    \caption{Cases 6 and 7 in the $\rho z$-plane: (a) Case 6 with $\mathcal{R}_1=1$, $\mathcal{R}_2=1.5$, $\mathcal{R}_3=2$ and (b) Case 7 with $\mathcal{R}_1=1$, $\mathcal{R}_2=1.5$, $\mathcal{R}_3=2$.}
    \label{F.case89}
\end{figure}

As Fig. \ref{F.case89} illustrates, Cases 6 and 7 feature two layers besides the borehole, where the mud layer can represent an invasion zone with a resistivity between those of the borehole and the outer formation.
\begin{table}[!htbp]
\begin{center}
\renewcommand{\arraystretch}{1.2}
\setlength{\tabcolsep}{14pt}
\caption{Comparison of electric potential for Case 6.}
    \begin{tabular}{|c|c|c|}
        \hline
         & FEM & Present Algorithm \\
        \hline
        $V_{16^{\prime\prime}}$ & 3.7097 $\times\;10^{-1}$ & 3.9061 $\times\;10^{-1}$ (2 sec.)\\
        $V_{32^{\prime\prime}}$ & 1.9796 $\times\;10^{-1}$ & 2.0245 $\times\;10^{-1}$ (2 sec.)\\
        \hline
        $\Delta V$              & 1.7301 $\times\;10^{-1}$ & 1.8816 $\times\;10^{-1}$\\
        \hline
    \end{tabular}
    \label{T.case8}
\vspace{4em}
\caption{Comparison of electric potential for Case 7.}
    \begin{tabular}{|c|c|c|}
        \hline
         & FEM & Present Algorithm \\
        \hline
        $V_{16^{\prime\prime}}$ & 3.6077 $\times\;10^{-1}$ & 3.8160 $\times\;10^{-1}$ (2 sec.)\\
        $V_{32^{\prime\prime}}$ & 1.9694 $\times\;10^{-1}$ & 2.0156 $\times\;10^{-1}$ (2 sec.)\\
        \hline
        $\Delta V$              & 1.6383 $\times\;10^{-1}$ & 1.8003 $\times\;10^{-1}$\\
        \hline
    \end{tabular}
    \label{T.case9}
\end{center}
\end{table}

\subsection{Electric potential maps}
\label{sec.4.2}
\begin{figure}[!htbp]
	\centering
    \subfloat[\label{F.case4.3D}]{%
      \includegraphics[width=2.8in]{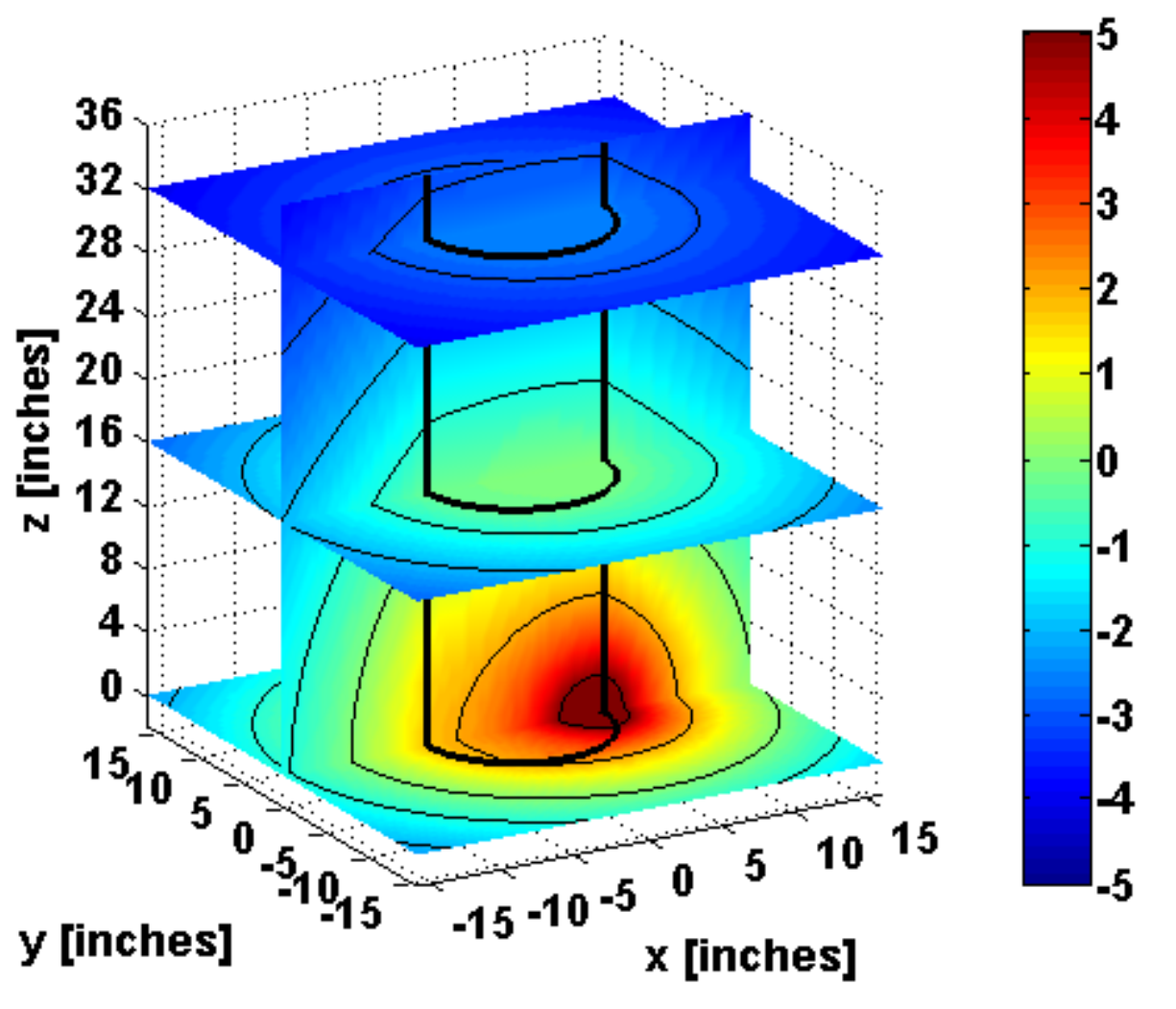}
    }
    \hspace{1.0cm}
    \subfloat[\label{F.case5.3D}]{%
      \includegraphics[width=2.8in]{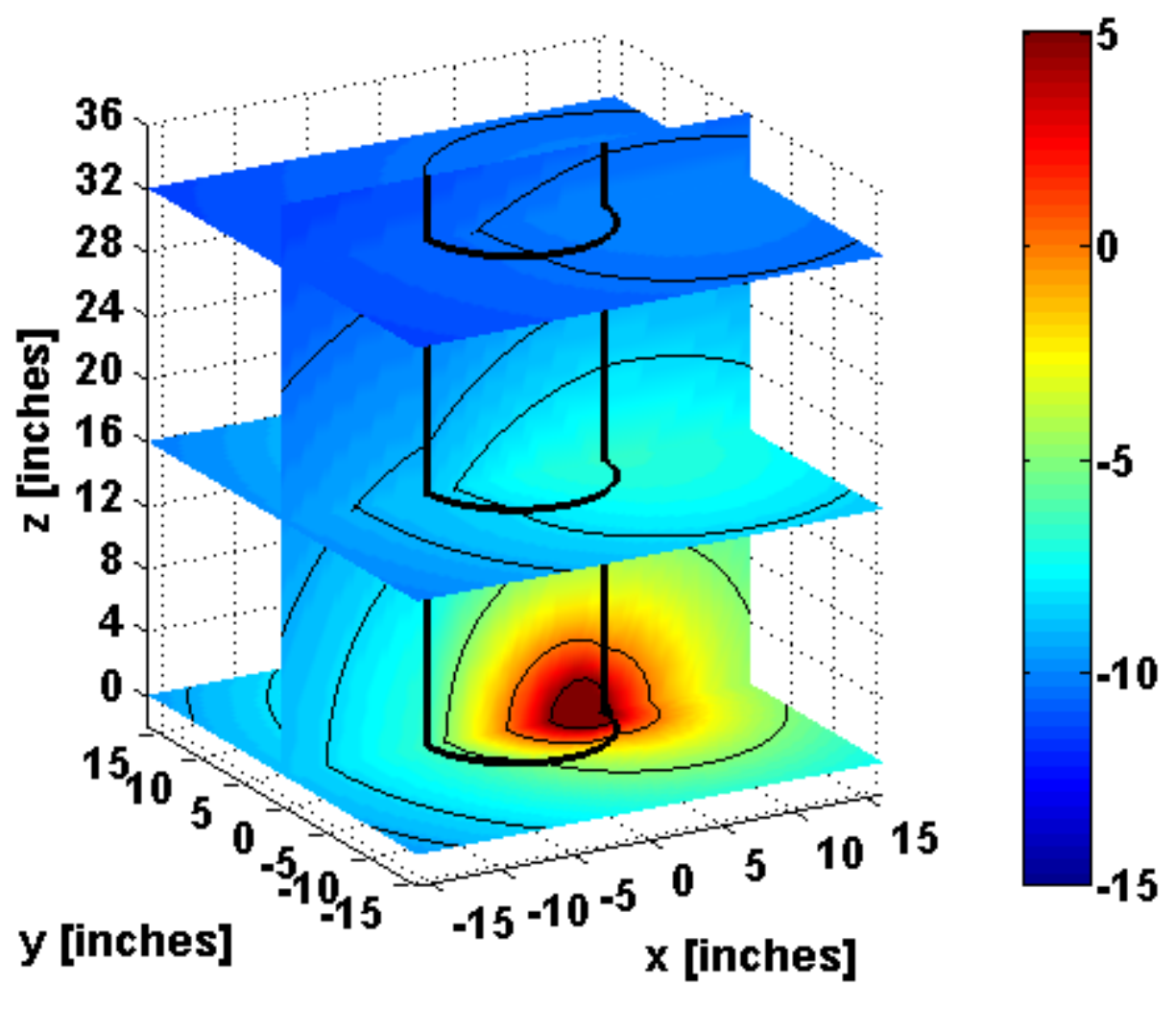}
    }\\
    \subfloat[\label{F.case6.3D}]{%
      \includegraphics[width=2.8in]{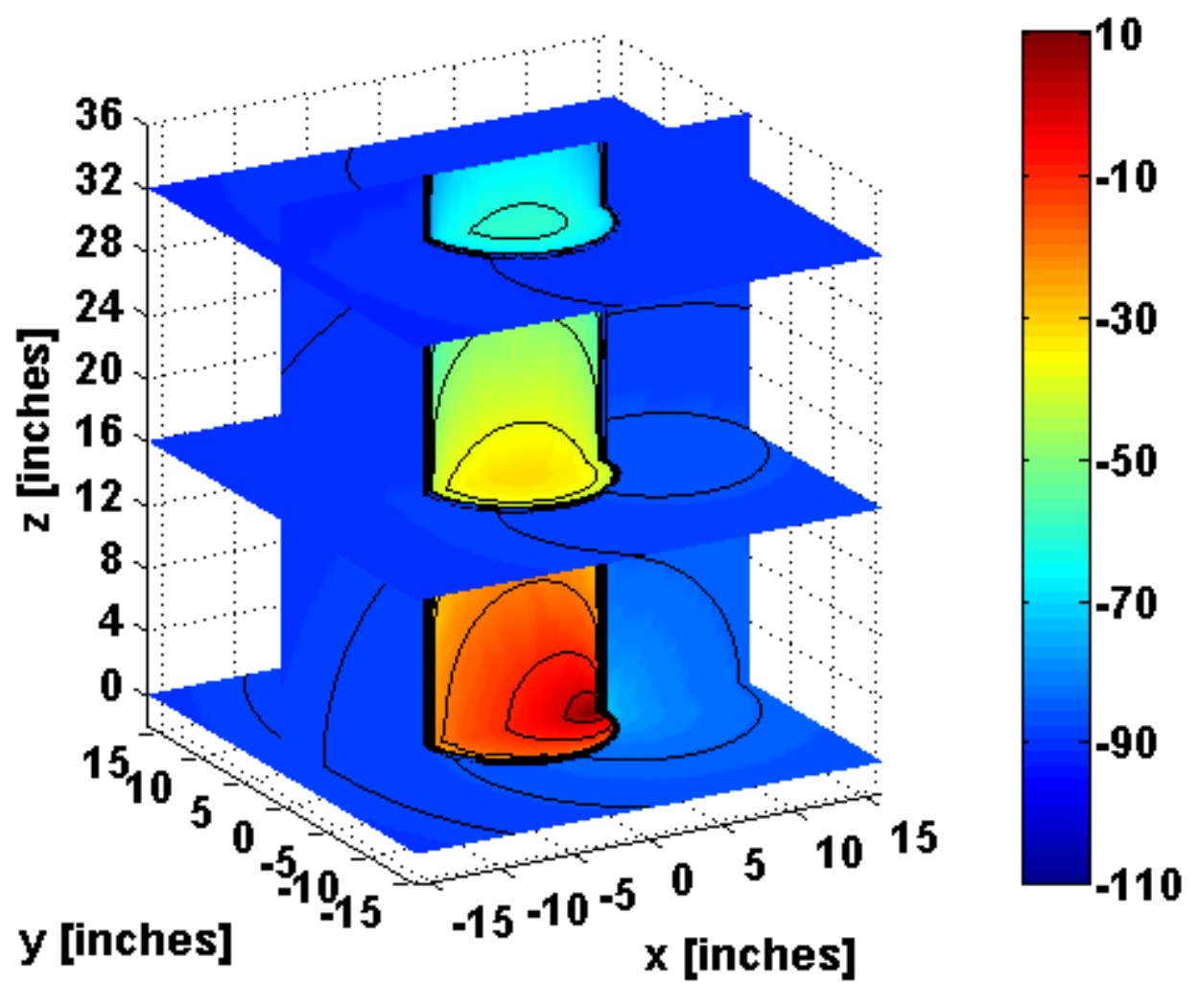}
    }
    \hspace{1.0cm}
    \subfloat[\label{F.case7.3D}]{%
      \includegraphics[width=2.8in]{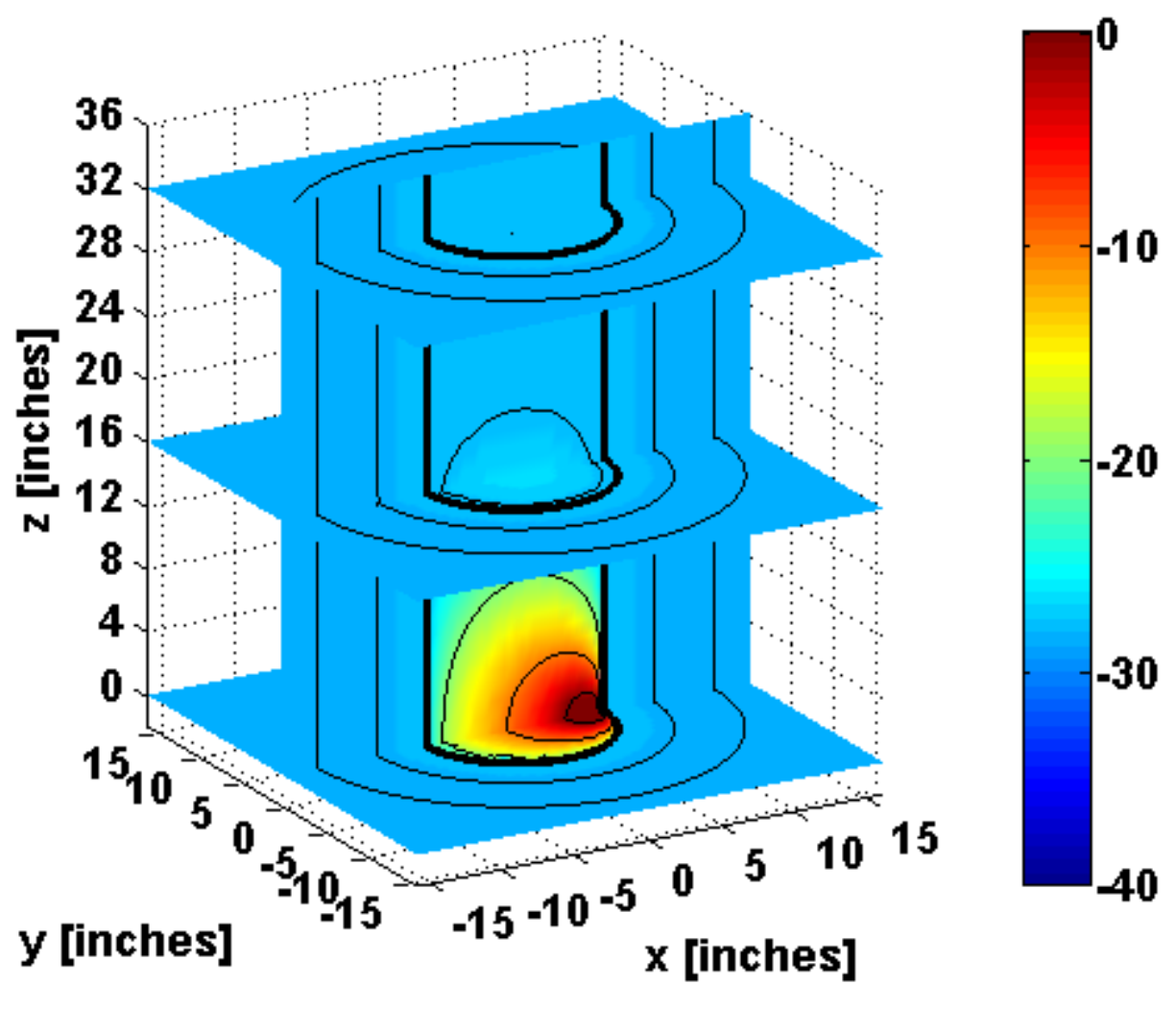}
    }
    \caption{Electric potential around the current electrode: (a) Case 2, (b) Case 3, (c) Case 4, and (d) Case 5.}
    \label{F.3D.plots}
\end{figure}

\begin{figure}[!htbp]
	\centering
    \subfloat[\label{F.case4.z16}]{%
      \includegraphics[width=2.8in]{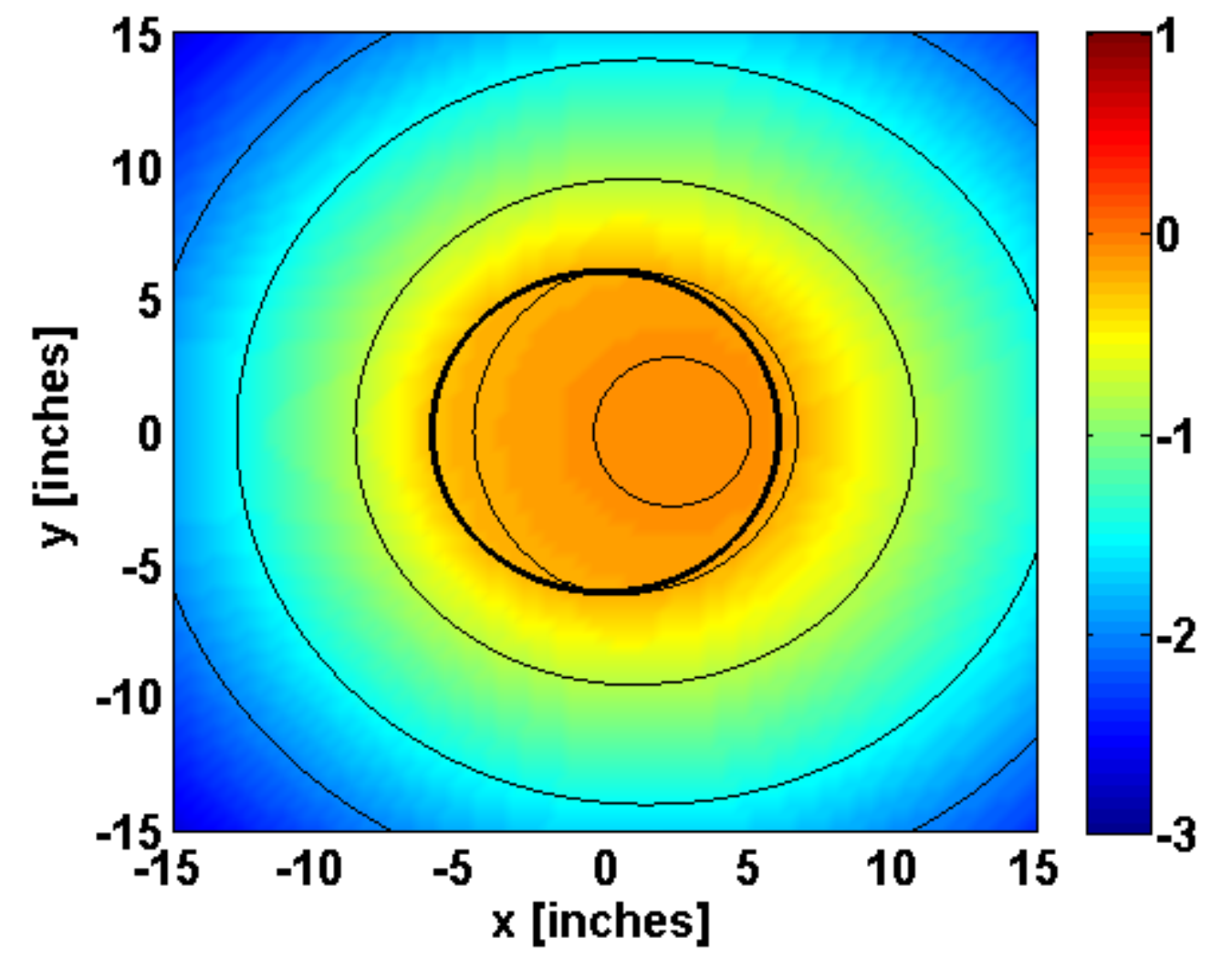}
    }
    \hspace{1.0cm}
    \subfloat[\label{F.case5.z16}]{%
      \includegraphics[width=2.8in]{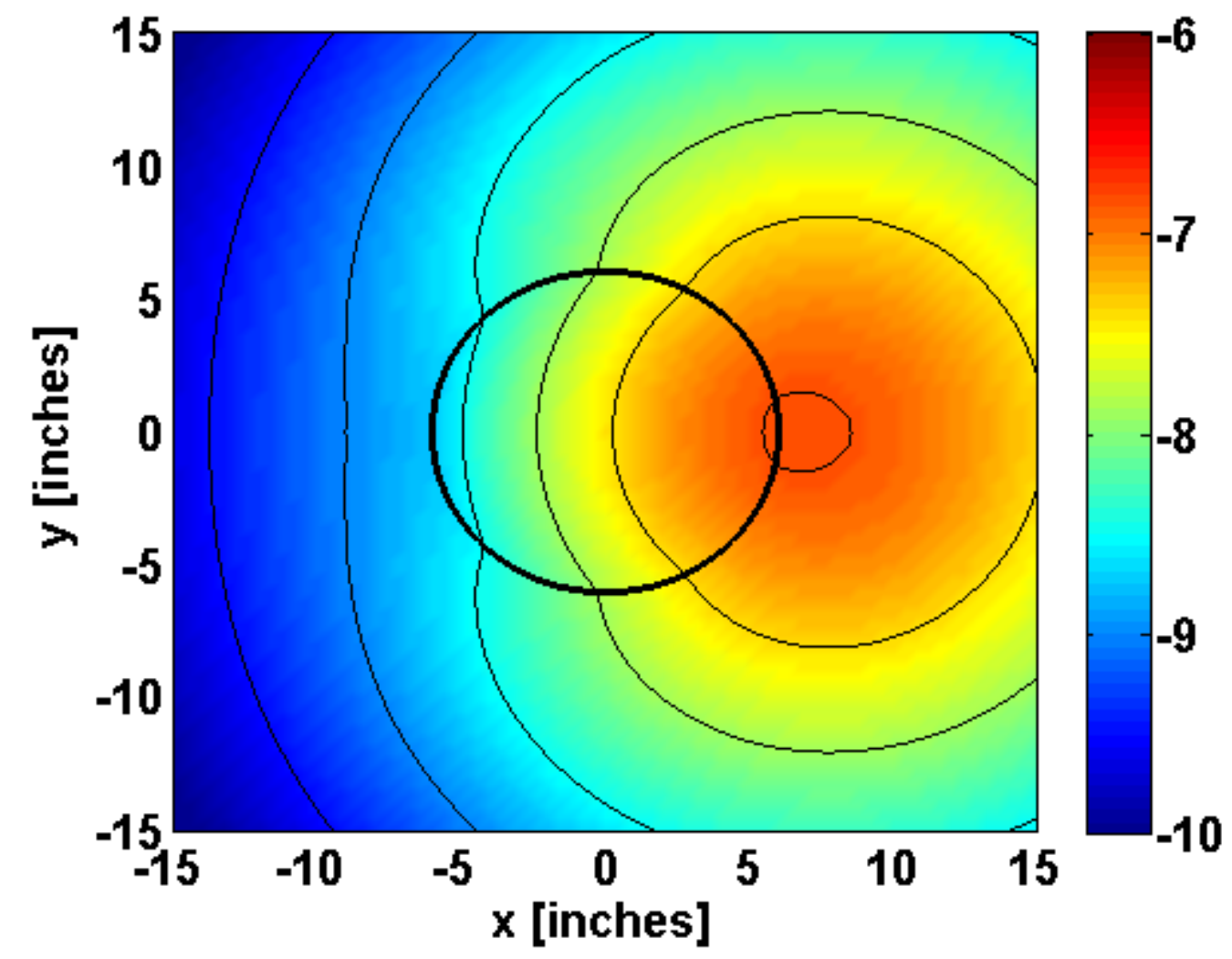}
    }\\
    \subfloat[\label{F.case6.z16}]{%
      \includegraphics[width=2.8in]{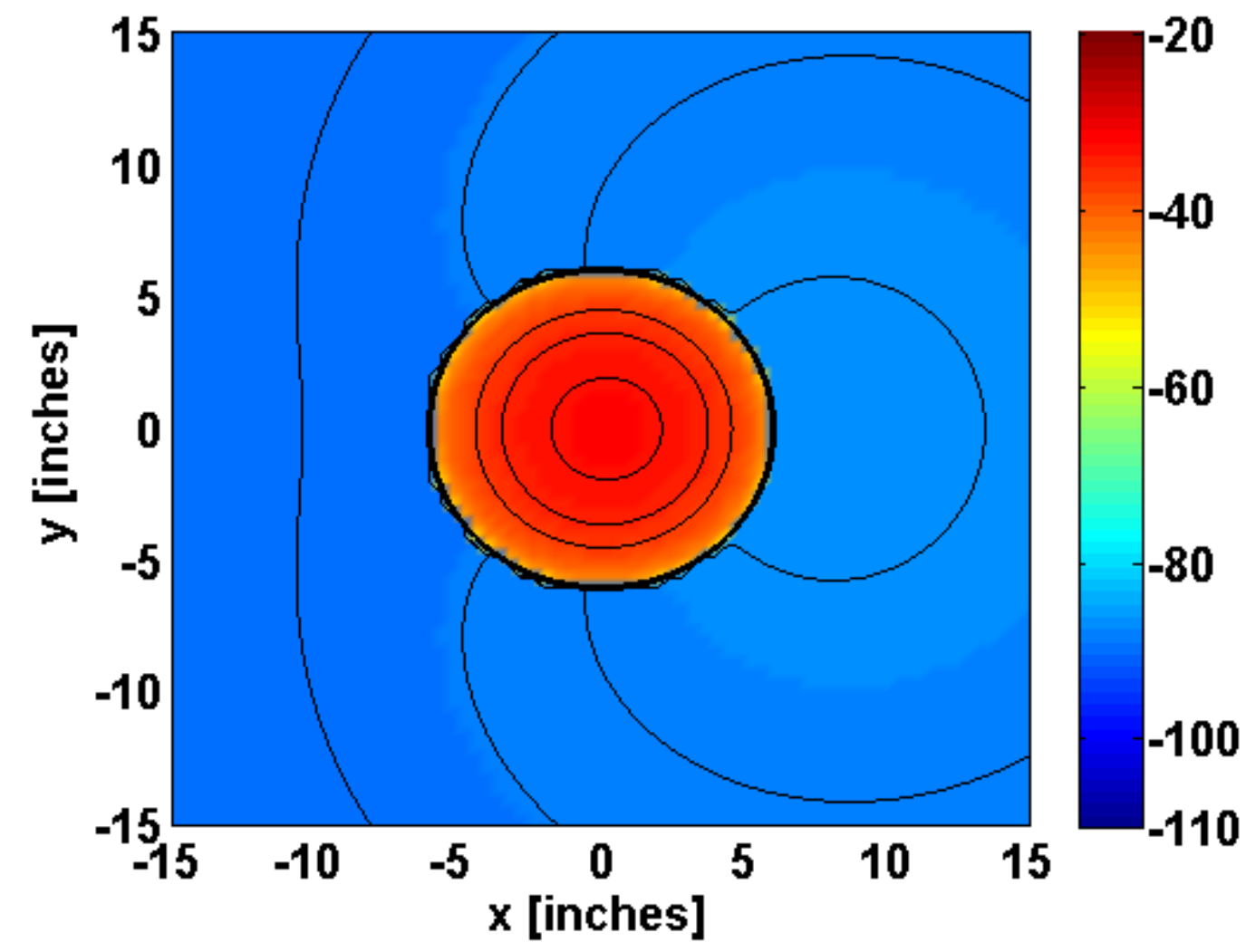}
    }
    \hspace{1.0cm}
    \subfloat[\label{F.case7.z16}]{%
      \includegraphics[width=2.8in]{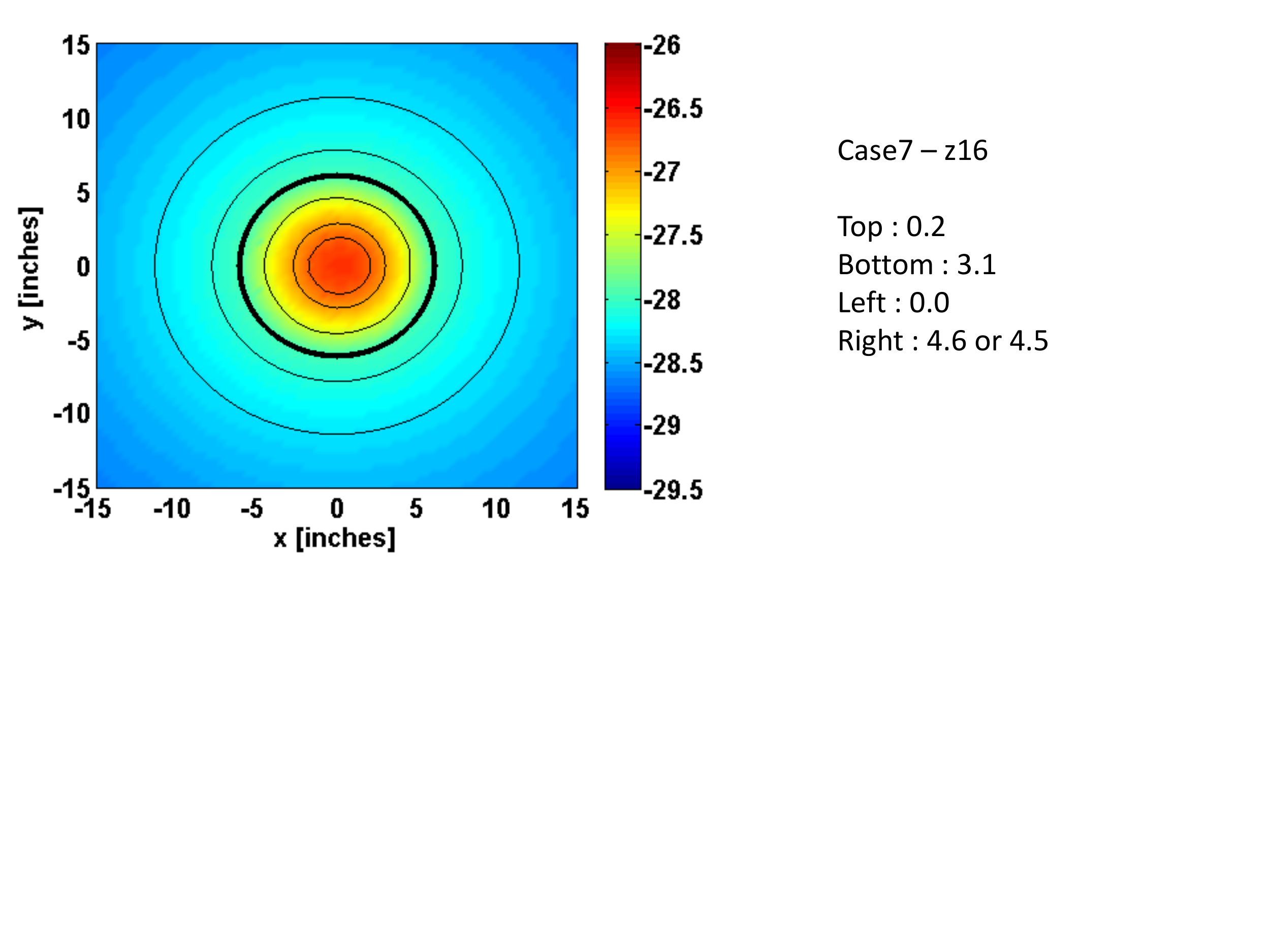}
    }
    \caption{Electric potential at $z=16^{\prime\prime}$ plane around the current electrode: (a) Case 2, (b) Case 3, (c) Case 4, and (d) Case 5.}
    \label{F.z16.plots}
\end{figure}

\begin{figure}[!htbp]
	\centering
    \subfloat[\label{F.case4.z32}]{%
      \includegraphics[width=2.8in]{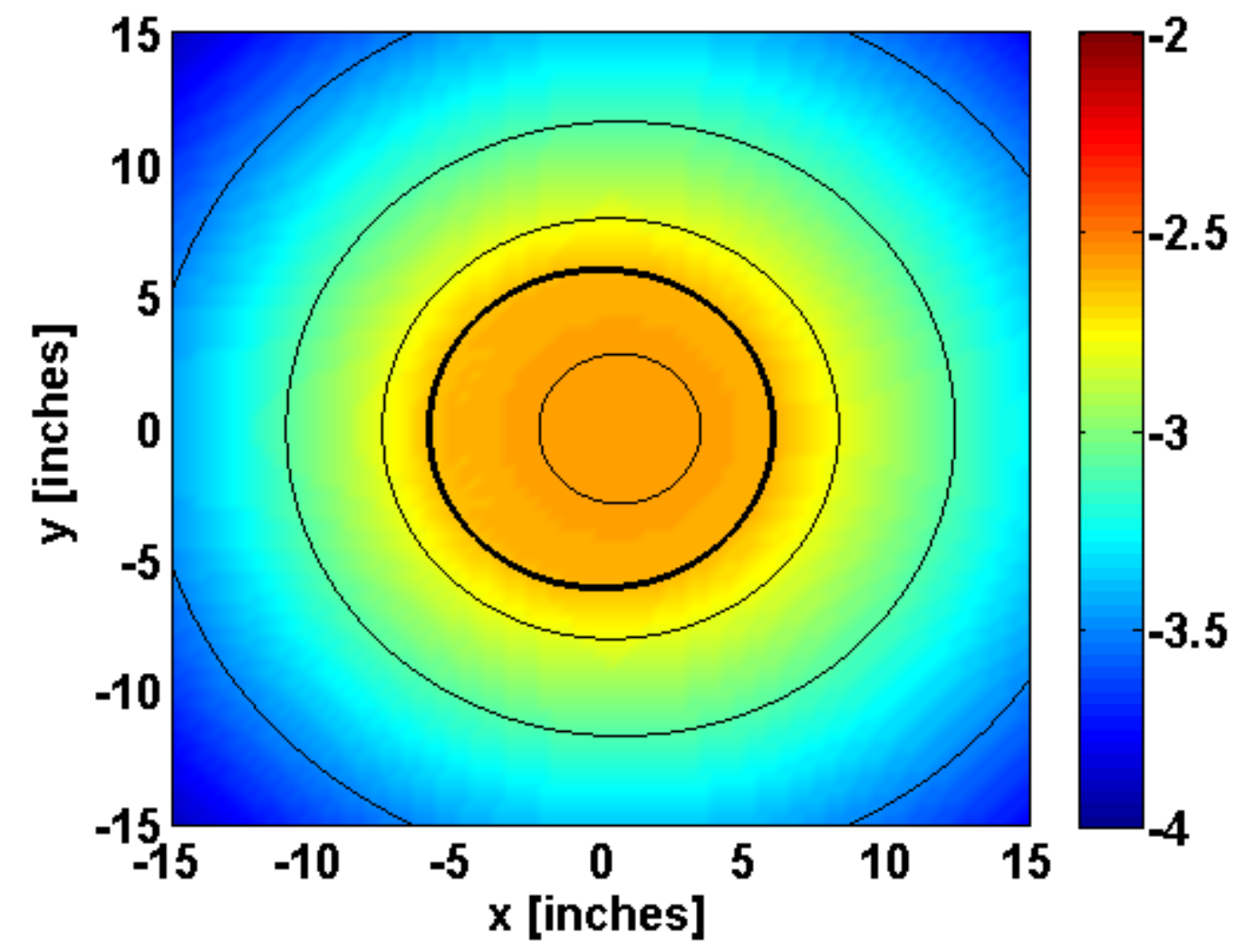}
    }
    \hspace{1.0cm}
    \subfloat[\label{F.case5.z32}]{%
      \includegraphics[width=2.8in]{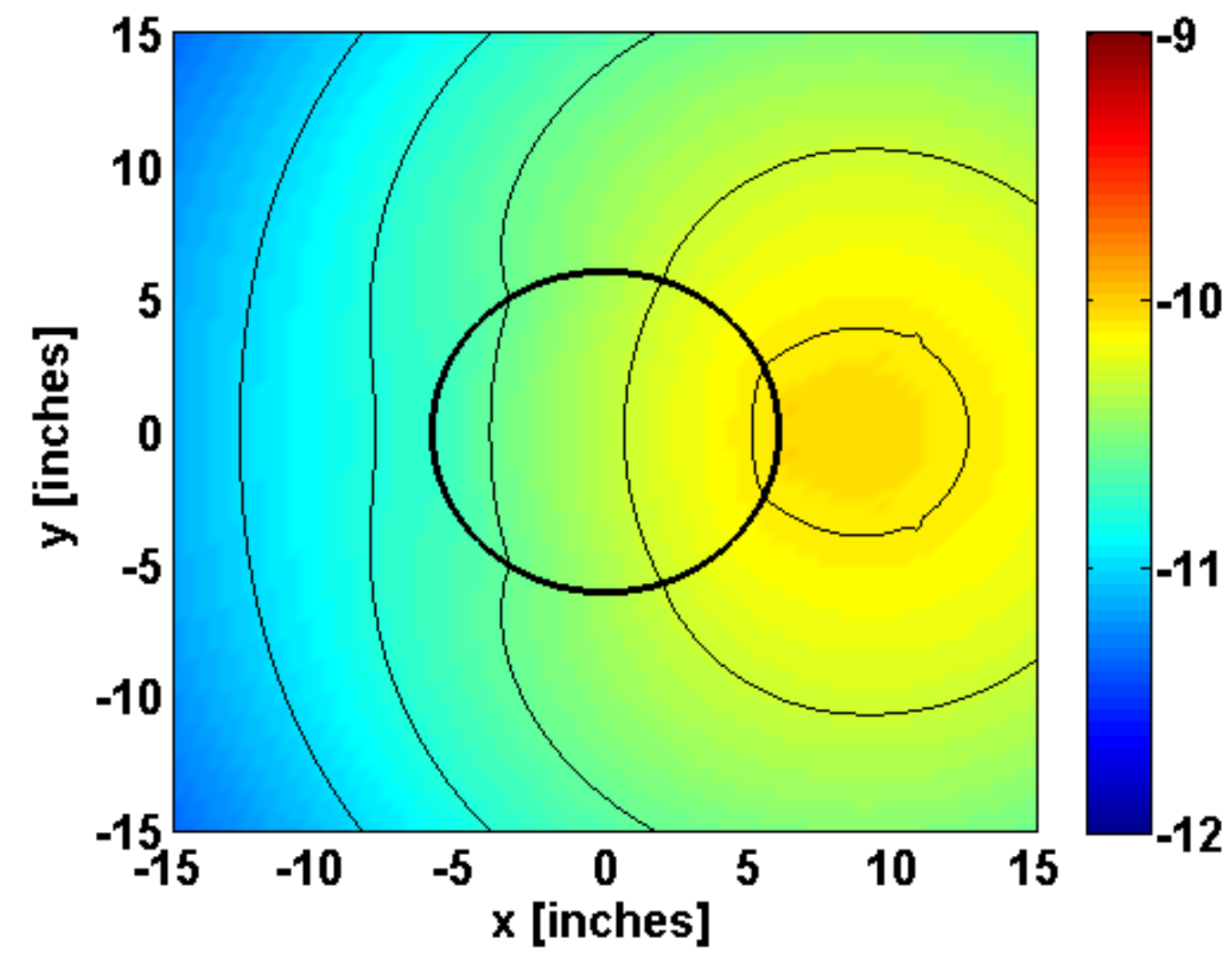}
    }\\
    \subfloat[\label{F.case6.z32}]{%
      \includegraphics[width=2.8in]{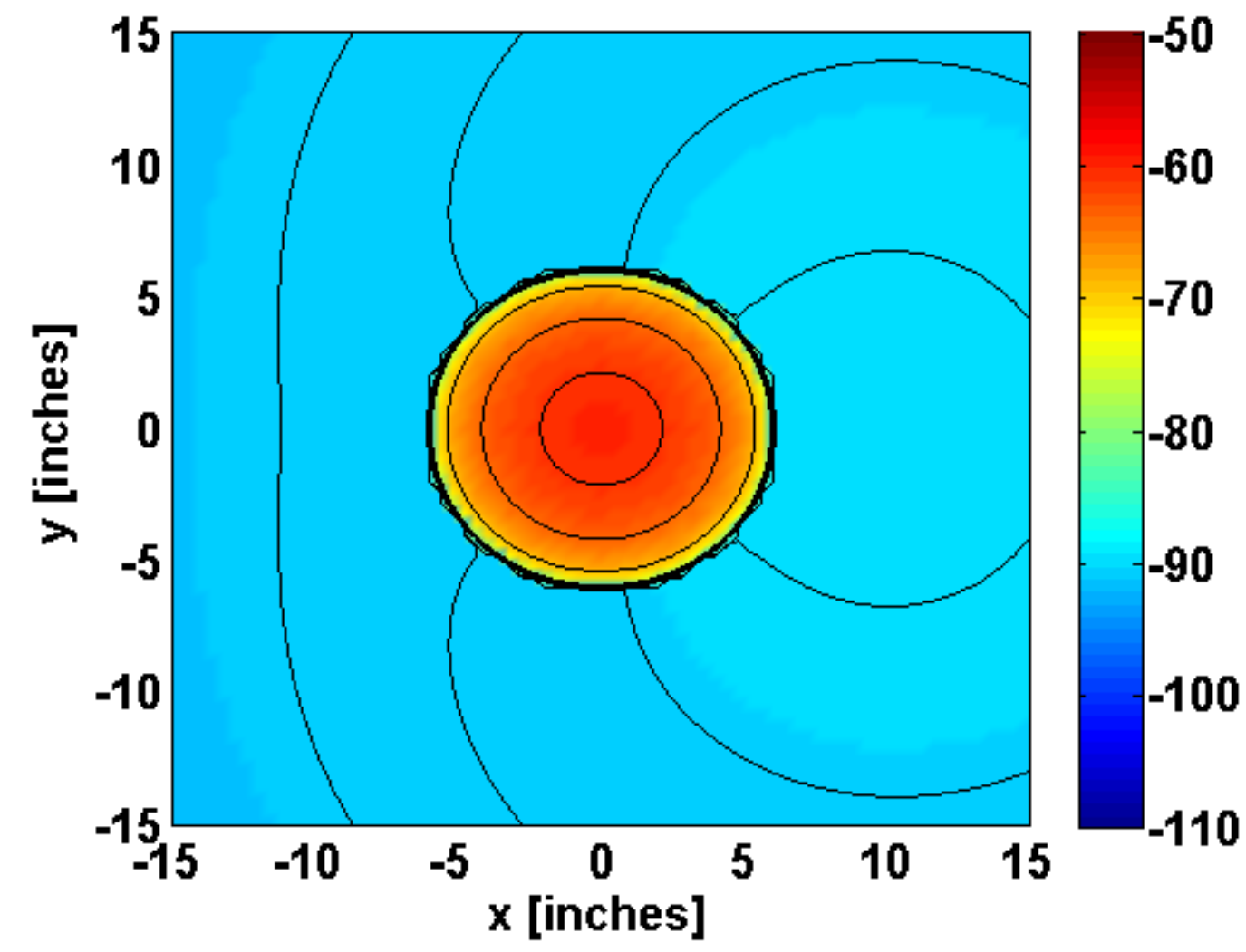}
    }
    \hspace{1.0cm}
    \subfloat[\label{F.case7.z32}]{%
      \includegraphics[width=2.8in]{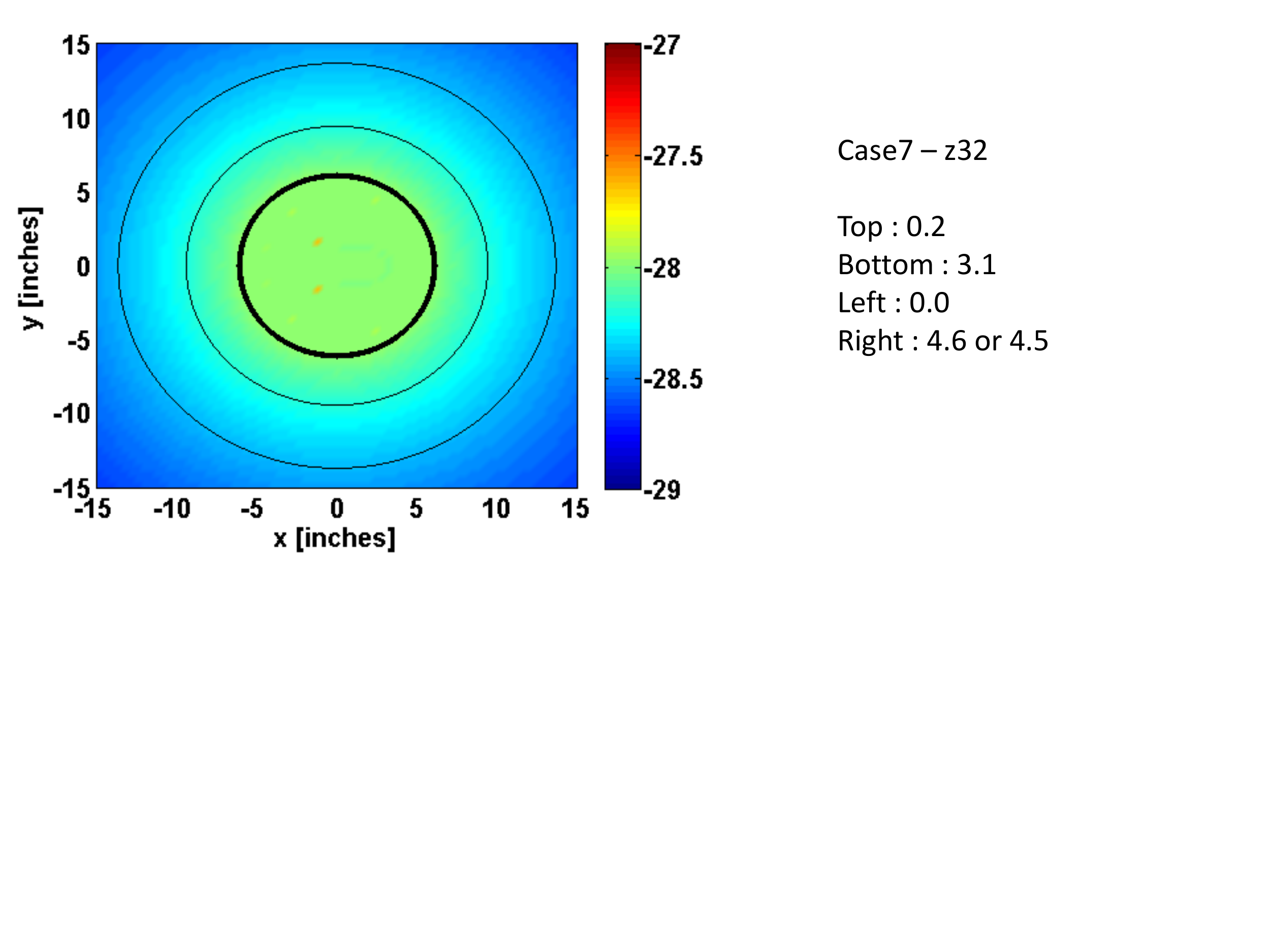}
    }
    \caption{Electric potential at $z=32^{\prime\prime}$ plane around the current electrode: (a) Case 2, (b) Case 3, (c) Case 4, and (d) Case 5.}
    \label{F.z32.plots}
\end{figure}

\begin{figure}[!htbp]
	\centering
    \subfloat[\label{F.case4.y0}]{%
      \includegraphics[width=2.8in]{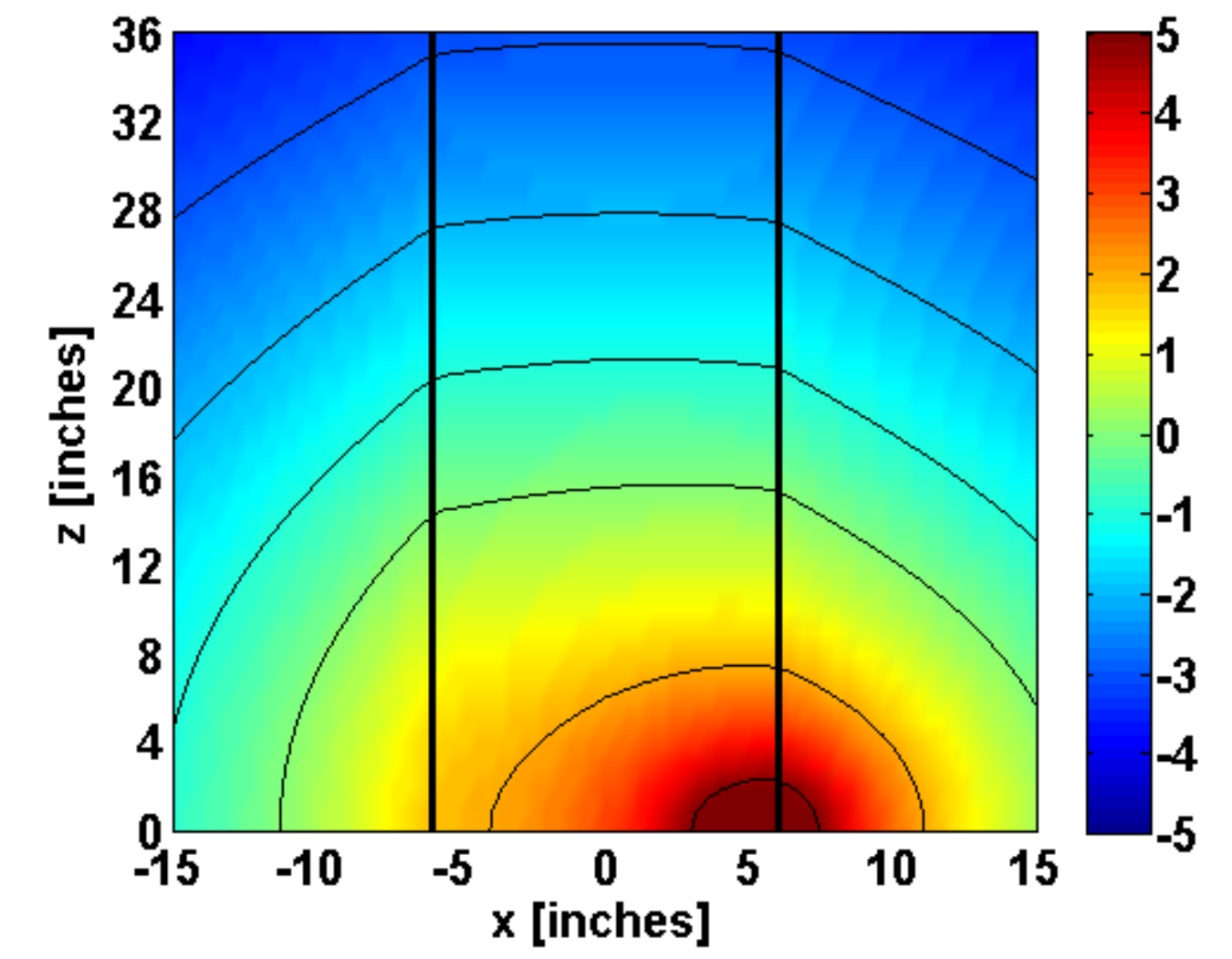}
    }
    \hspace{1.0cm}
    \subfloat[\label{F.case5.y0}]{%
      \includegraphics[width=2.8in]{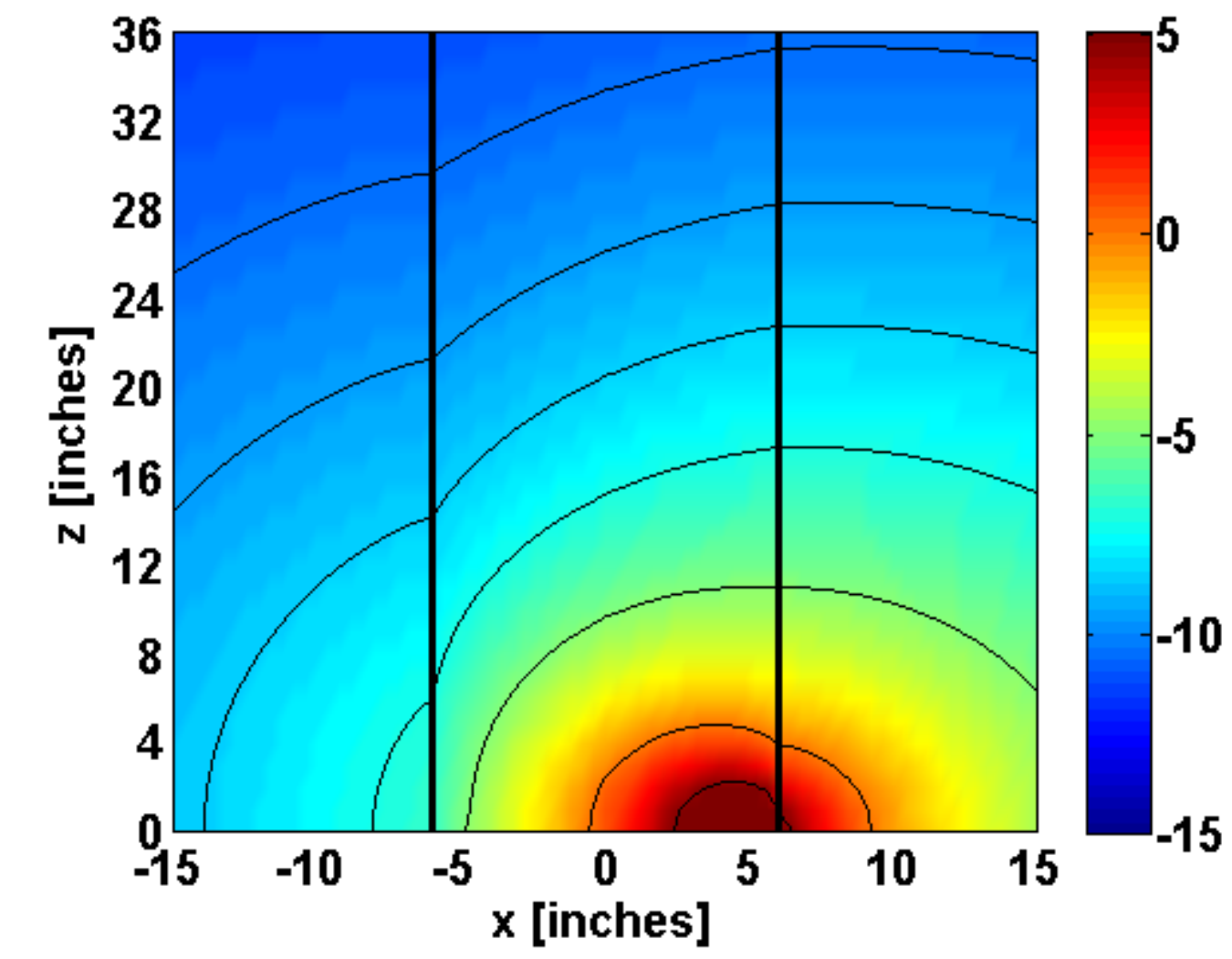}
    }\\
    \subfloat[\label{F.case6.y0}]{%
      \includegraphics[width=2.8in]{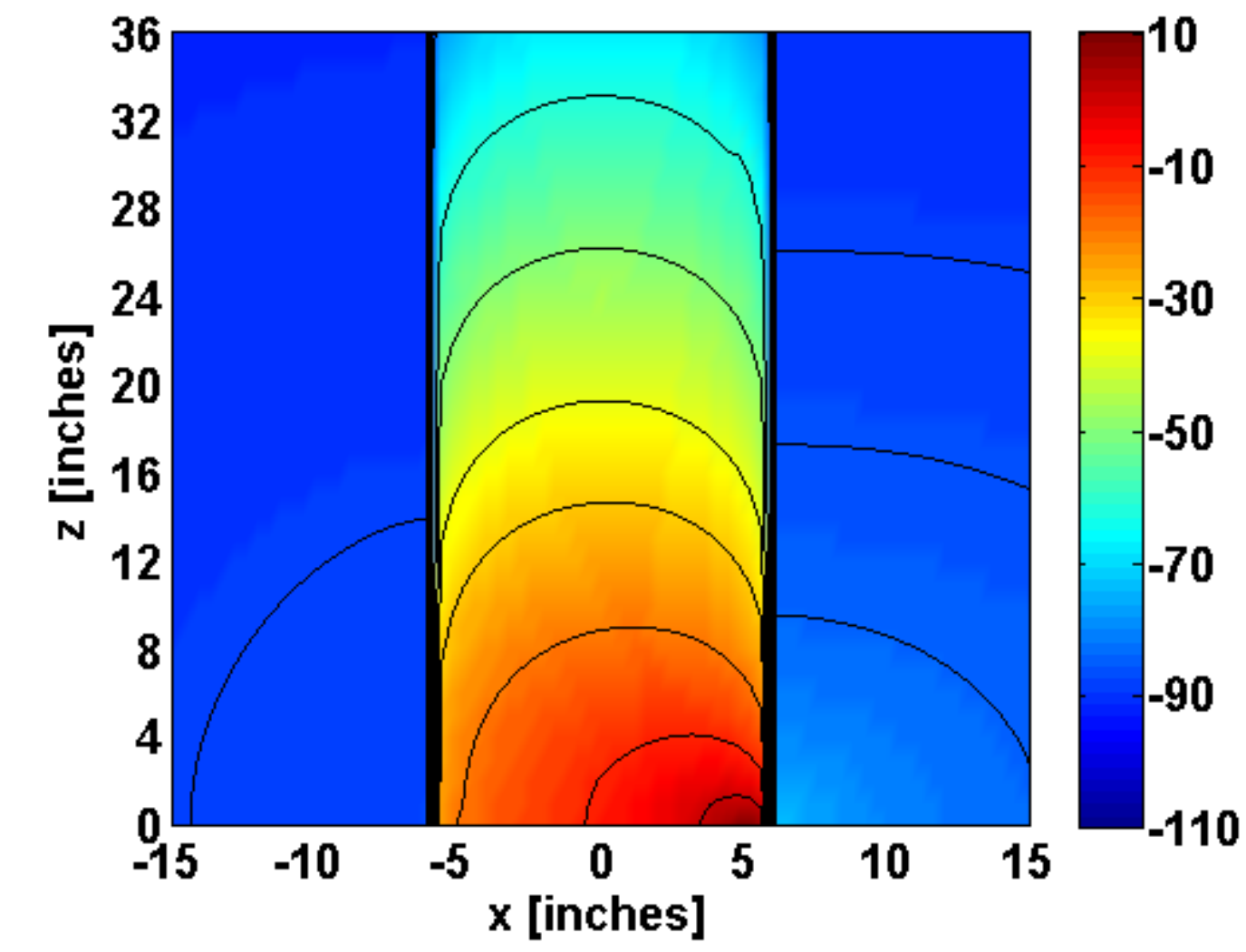}
    }
    \hspace{1.0cm}
    \subfloat[\label{F.case7.y0}]{%
      \includegraphics[width=2.8in]{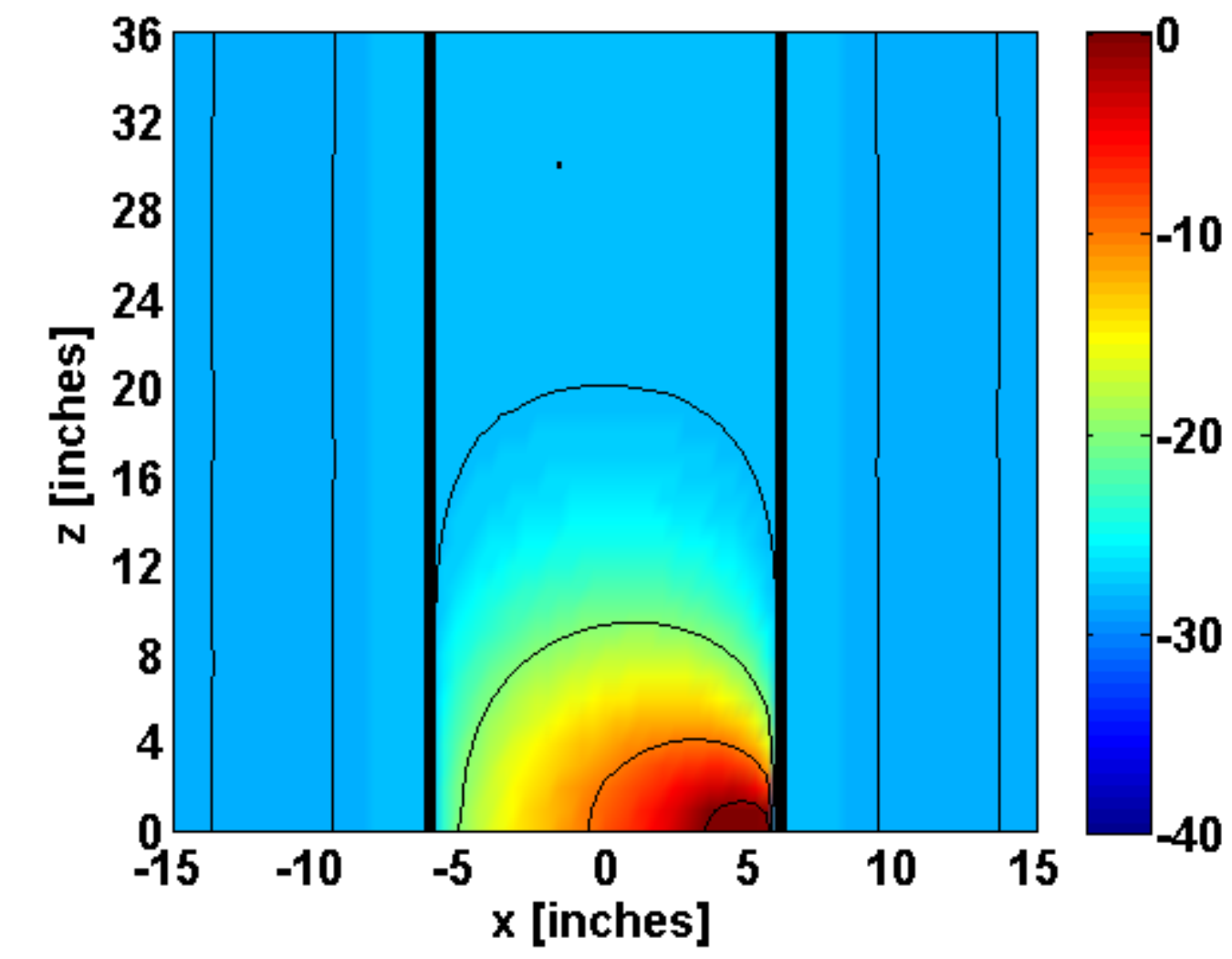}
    }
    \caption{Electric potential at $y=0^{\prime\prime}$ plane around the current electrode: (a) Case 2, (b) Case 3, (c) Case 4, and (d) Case 5.}
    \label{F.y0.plots}
\end{figure}

Plots of the spatial distributions of the electric potential of Cases 2--5 are provided next. Since the potential varies by many orders of magnitude near the electrodes, a log-scale is used; i.e. we plot the quantity $10\log_{10}|V|$. Fig. \ref{F.3D.plots} depicts the potential distributions using a three-dimensional view. Fig. \ref{F.z16.plots} and Fig. \ref{F.z32.plots} show the cross-sections of the potential distributions at the $z=16^{\prime\prime}$ and $z=32^{\prime\prime}$ planes, respectively, while Fig. \ref{F.y0.plots} shows the cross-sections at the $y=0^{\prime\prime}$ plane. In all figures, thicker black lines represent interfaces between cylindrical layers and thinner black lines represent equipotential contours. Note that the third layer of Case 5 is too thin for visualization in these plots.

\section{Conclusion}
We have introduced a stable semianalytical formulation to compute the potential due to arbitrary-located electrodes in a cylindrically layered media, where the physical parameters (viz., layer resistivity) can vary by many order of magnitude.
Stability was achieved by a rescaling of the modified Bessel functions and subsequent manipulation of the integrand expressions to avoid underflow and overflow problems under double-precision arithmetic.
Several extrapolation methods to compute the involved Sommerfeld integrals were also considered and evaluated.
The resulting algorithm was verified in a number of cases relevant to borehole geophysics. The fast speed and robustness of the algorithm makes it also quite suited as a forward solver engine for the inverse roblem, where the formation
resistivity needs to be estimated from a limited numberof computed (measured) potential data. The same basic algorithm can be applied to other steady-state diffusion problems obeying Poisson's equation with discontinous coefficients in cylindrically layered geometries.

\section*{Acknowledgments}
The authors are grateful to Halliburton Energy Services for the permission to publish this work and to Dr. Baris Guner for generating validation data.

\section*{Appendix A: Folding azimuth summation}

The computation of potential in \eqref{psi.final} involves infinite series over the azimuth mode $n$.
For an arbitrary order $\nu\ge 0$ \cite{Jin:SpecialFunctions}, we have
\begin{subequations}
\begin{flalign}
I_{-\nu}(z)&=I_{\nu}(z)+\frac{2}{\pi}\sin(\nu\pi)K_{\nu}(z), \label{In.neg}\\
K_{-\nu}(z)&=K_{\nu}(z). \label{Kn.neg}
\end{flalign}
\end{subequations}
If the order is a positive integer, $\nu = n$, the above reduces to
$I_{-n}(z)=I_{n}(z)$ and $K_{-n}(z)=K_{n}(z)$. Consequently, we can write
\begin{flalign}
\psi_i
&=\frac{\mathcal{I}}{2\pi^2\sigma_j}\suma e^{\iu n(\phi-\phi')}\intzi
	F_n(\rho,\rho')\cos(\lam (z-z'))d\lam \notag\\
&=\frac{\mathcal{I}}{2\pi^2\sigma_j}\intzi\left[F_0(\rho,\rho')+2\sumb F_n(\rho,\rho')\cos(n(\phi-\phi'))\right]
    \cos(\lam (z-z'))d\lam. \label{psi.folding.azi}
\end{flalign}

\section*{Appendix B: Extrapolation methods and convergence study}
Among numerous extrapolation methods, some popular ones for Sommerfeld-type integrals are very briefly revisited here. For more details, the reader can refer, e.g., to \cite{Michalski98:Extrapolation, Weniger89:Nonlinear}. Before a given sequence is extrapolated, a Sommerfeld-type integral can be divided into a number of subintervals as
\begin{flalign}
S=\int_{a}^{\infty}g(\lam)p(\lam)d\lam
 =\sum_{i=0}^{\infty}\int_{\lam_{i-1}}^{\lam_i}g(\lam)p(\lam)d\lam
 =\sum_{i=0}^{\infty}u_i, \label{Sum}
\end{flalign}
where $g(\lam)$ is an exponentially decaying part and $p(\lam)$ is an oscillatory part. This is called the partition-extrapolation approach~\cite{Squire75:Partition}. In general, the remainders are defined as
\begin{flalign}
r_n=S_n-S=-\int_{\lam_n}^{\infty}g(\lam)p(\lam)d\lam. \label{rn}
\end{flalign}
Furthermore, the remainders are assumed to feature Poincar\'{e}-type asymptotic expansions \cite{Michalski98:Extrapolation} of the form
\begin{flalign}
r_n\sim \omega_n\sum_{i=0}^{\infty}a_i\lam^{-i}_n,\qquad n\rightarrow\infty, \label{rn.exp}
\end{flalign}
where $\omega_n$ is the remainder estimate and $a_i$ are associated coefficients. The estimates $\omega_n$ play an important role in the extrapolation and can be obtained analytically or numerically. The coefficients $a_i$ are unknowns but they are not necessary for the extrapolation itself. In our case, $g(\lam)$ in \eqref{Sum} can be asymptotically expressed as
\begin{flalign}
g(\lam)\sim\frac{e^{-\lam|\rho-\rho'|}}{\lam}\sum_{i=0}^{\infty}\frac{c_i}{\lam^{i}},
	\label{g.asymtotic}
\end{flalign}
where $c_i$ are arbitrary constants. Furthermore, $p(\lam)= \cos(\lam(z-z'))$ with half-period equal to $\pi/|\rho-\rho'|$. After some algebra, it can be shown that the remainder estimates write as
\begin{flalign}
\omega_n=\frac{(-1)^{n+1}}{\lam_n}e^{-\frac{n\pi|\rho-\rho'|}{|z-z'|}}. \label{wn}
\end{flalign}
We list next three popular extrapolation methods for a given sequence, $\{S_n\}=S_{0},S_{1},S_{2},\cdots,S_{n}$, where we consider $S_{n}=S_{n}^{(0)}$.

\vskip 0.1in
\noindent {\it (i) Euler transformation}
\begin{flalign}
S_n^{(k+1)}=\frac{1}{2}\left(S_n^{(k)}+S_{n+1}^{(k)}\right),\qquad n,k\geq 0 \label{Euler}
\end{flalign}
The best approximation in this case is $S_0^{(k)}|_{k=n}$, and this choice is most effective for logarithmic alternating sequences.

\vskip 0.1in
\noindent {\it (ii) Iterative Aitken transformation}
\begin{flalign}
S_n^{(k+1)}=S_n^{(k)} - \frac{[\Delta S_n^{(k)}]^2}{\Delta^2 S_n^{(k)}},\qquad n,k\geq 0
	\label{Aitken}
\end{flalign}
Note that $\Delta S^{(k)}_n=S_{n+1}^{(k)}-S_n^{(k)}$ and $\Delta^2 S_n^{(k)}=S_{n+2}^{(k)}-2S_{n+1}^{(k)}+S_n^{(k)}$.
Obviously, this is a nonlinear transformation and it can be applied to both linear monotone and alternating sequences. When an odd number of sequences is given, the best approximation is $S_0^{(k)}|_{k=n}$. For an even number, the best approximation is $S_1^{(k)}|_{k=n}$.

\vskip 0.1in
\noindent {\it (iii) Weighted-averages method}
\begin{flalign}
S^{(k+1)}_n=\frac{S^{(k)}_n+\eta^{(k)}_n S^{(k)}_{n+1}}{1+\eta^{(k)}_n},\qquad n,k\geq 0,
	\label{WAM}
\end{flalign}
where $\eta^{(k)}$ is the weight and defined as
\begin{flalign}
\eta^{(k)}=-\frac{\omega_n}{\omega_{n+1}}
		  =\frac{\lam_{n+1}}{\lam_n}e^{\frac{\pi|\rho-\rho'|}{|z-z'|}}.
	\label{eta}
\end{flalign}
The weighted-averages method \cite{Mosig83:Analytical,Mosig12:Weighted} can be regarded as generalized Euler transformation since $\eta=1$ recovers the Euler transformation. The relative
power of this method, compared to the two other methods, comes from using remainder estimates. As pointed out in \cite{Michalski98:Extrapolation}, the weighted-averages method is quited suited for extrapolating Sommerfeld-type integrals.

\begin{figure}[t]
	\centering
	\subfloat[\label{F.type1}]{%
      \includegraphics[width=2.7in]{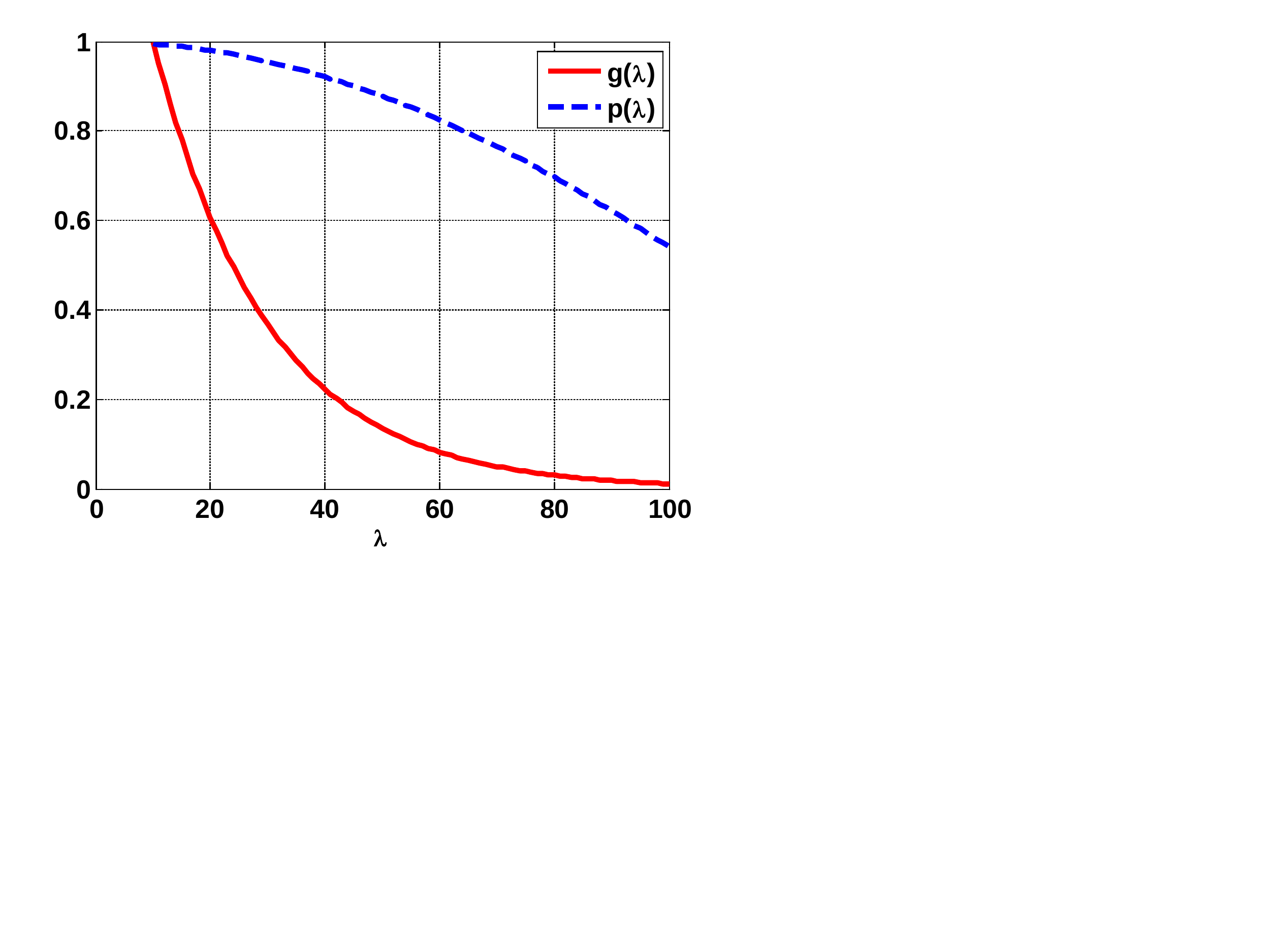}
    }
    \hspace{1.0cm}
    \subfloat[\label{F.type2}]{%
      \includegraphics[width=2.7in]{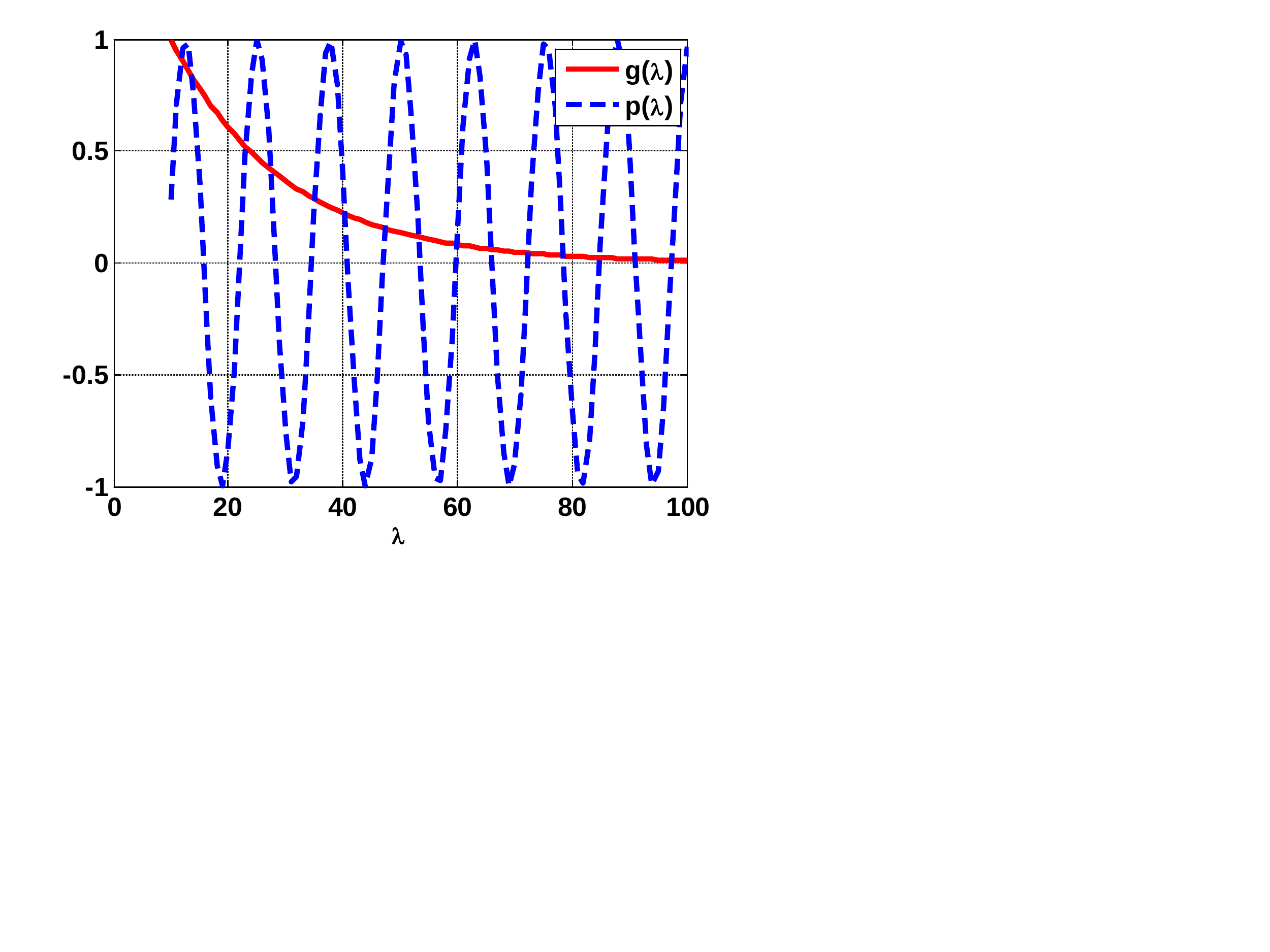}
    }
    \caption{Two scenarios of the integrand: (a) Type 1 with $|\rho-\rho'|=0.05$, $|z-z'|=0.01$ and (b) Type 2 with $|\rho-\rho'|=0.05$, $|z-z'|=0.5$.}
    \label{F.two.types}
\end{figure}

\begin{figure}[t]
    \centering
    \includegraphics[width=2.5in]{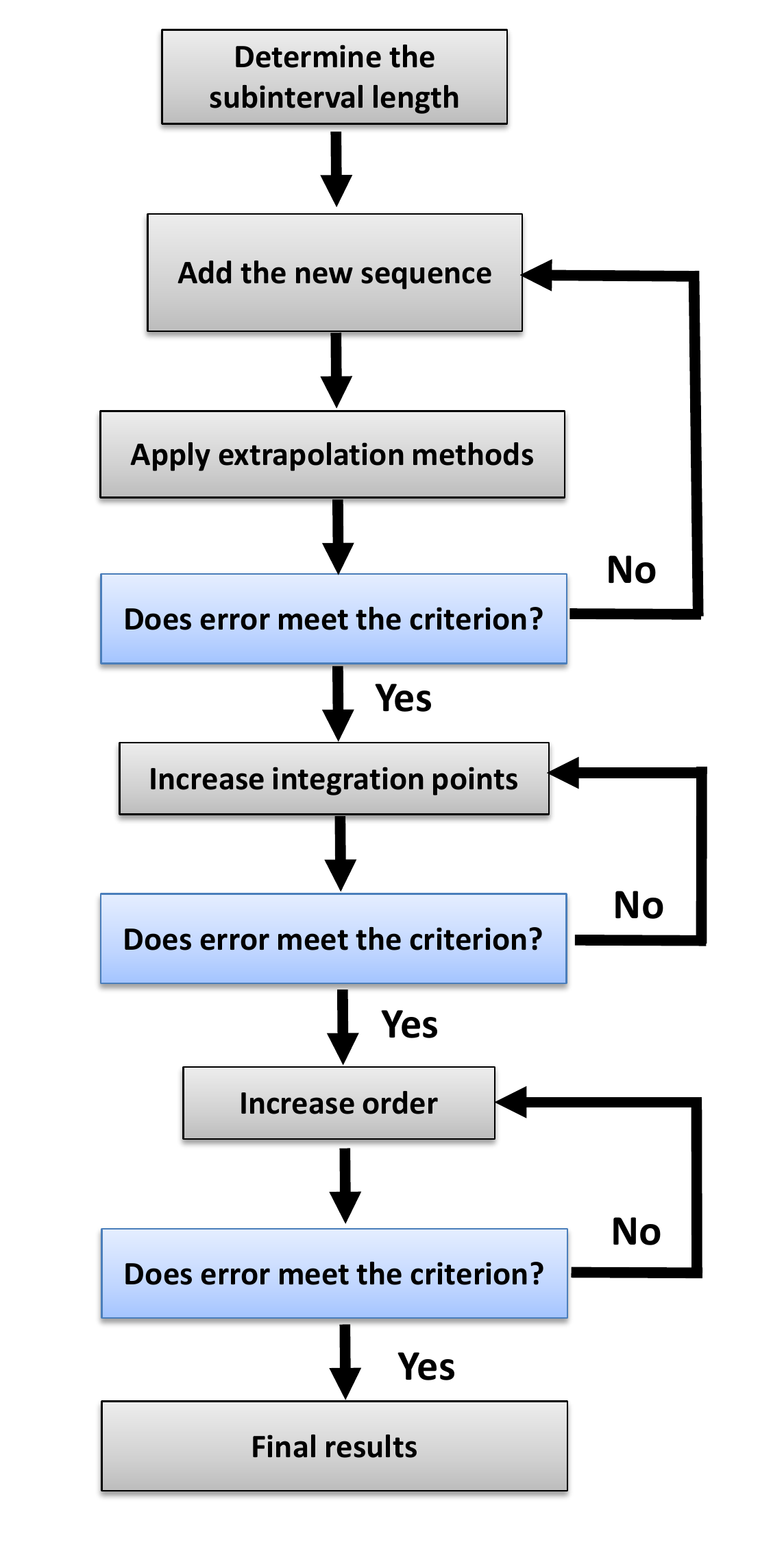}\\
    \caption{Flowchart of the procedure for computation of electric potential using extrapolation methods.}
    \label{F.procedure}
\end{figure}

\vskip 0.1in
To evaluate the integral \eqref{Sum}, the subinterval length $q$ should be first determined. As suggested in \cite{Michalski98:Extrapolation}, the half-period of the oscillating part of the integrand is a good choice for $q$ because it makes the sequences alternating; i.e.,
$q=\pi/|z-z'|$.
However, this is not appropriate here in some circumstances. Let us consider the two scenarios depicted in Fig. \ref{F.two.types}. The two figures show the behavior of $g(\lam)$ and $p(\lam)$ for different combinations of $|\rho-\rho'|$ and $|z-z'|$ as $\lam$ increases. For better visualization, the functions are normalized by their respective maxima.
Type 1 integrand occurs when $|\rho-\rho'|>|z-z'|$ and shows linear monotone convergence. On the other hand, Type 2 integrand occurs when $|\rho-\rho'|<|z-z'|$ and shows logarithmic alternating convergence. Therefore, such an expression for $q$ is not appropriate for Type 1 because the integrand is near zero before the first half-period comes. For the Type 1 integrand, a different subinterval length can be defined as
$q=\pi/|\rho-\rho'|$, so that the subinterval length can be written in general as
\begin{flalign}
q=\frac{\pi}{\text{max}(|\rho-\rho'|,|z-z'|)}. \label{q.both}
\end{flalign}
Once $q$ is determined, it becomes necessary to determine how many subintervals are required to reach sufficient convergence. To do so, the relative error below is defined
\begin{flalign}
e_i=\frac{|T(S_{i+1})-T(S_i)|}{|T(S_{i+1})|},
\end{flalign}
where $T(S_i)$ is the extrapolated (transformed) value for given sequences, $\{S_i\}=S_{0},S_{1},\cdots,S_i$. If $e_i$ is less than a given error tolerance $e_{tol}$, the sequence is stopped and the number of subintervals is determined. Next, the number of quadrature points per the subinterval is increased until the
relative error between two adjacent iterations meets the desired criterion. To distinguish it from $e_{tol}$, the criterion in this step is denoted error threshold $e_{thr}$. The same step is repeated for determining the number of orders. The overall procedure is schematically depicted in Fig. \ref{F.procedure}.

Some convergence tests are performed next to evaluate the numerical integration procedures above. The domain is assumed homogeneous because exact (analytical) solutions are available as reference. Three cases of source/observation distances are considered.
\begin{center}
Case 1 : $\;|\rho-\rho'| = 0.001\text{ m}$, $\;|z-z'| = 0.1\text{ m}$\\
Case 2 : $\;|\rho-\rho'| = \quad0.1\text{ m}$, $\;|z-z'| = 0.1\text{ m}$\\
Case 3 : $\;|\rho-\rho'| = \quad\;10\text{ m}$, $\;|z-z'| = 0.1\text{ m}$\\
\end{center}
In each case, the three extrapolation methods mentioned before are compared. The excitation current magnitude and the medium resistivity are both set to one. The relevant error parameters are $e_{tol}=10^{-6}$ and $e_{thr}=10^{-4}$. The smaller $e_{tol}$ is chosen to examine the effect of the methods on the number of subintervals. Tables \ref{T.p1}, \ref{T.p2}, and \ref{T.p3} compare the results, where the first rows represent the number of subintervals needed to achieve convergence in terms of $e_{tol}$ and the second rows represent the relative error against the analytical solution. As Table \ref{T.p1} shows, the iterative Aitken method does not work for Case 1 because it corresponds to a Type 2 integrand with logarithmic alternating convergence. The Euler transformation works well for all cases, but it can be seen that the weighted-averages method provides best results, corroborating the conclusions stated in~\cite{Michalski98:Extrapolation}.

\begin{table}[!htbp]
\begin{center}
\renewcommand{\arraystretch}{1.2}
\setlength{\tabcolsep}{12pt}
\caption{Comparison of the three extrapolation methods for Case 1.}
    \begin{tabular}{|c|c|c|c|}
        \hline
         & Euler & Aitken & weighted-averages \\
        \hline
        \# of subintervals & 16 & $>$100 & 10\\
		Relative error & 1.9990 $\times\;10^{-6}$ & N. A. & 3.5863 $\times\;10^{-7}$\\
        \hline
    \end{tabular}
    \label{T.p1}
\end{center}
\end{table}

\begin{table}[!htbp]
\begin{center}
\renewcommand{\arraystretch}{1.2}
\setlength{\tabcolsep}{12pt}
\caption{Comparison of the three extrapolation methods for Case 2.}
    \begin{tabular}{|c|c|c|c|}
        \hline
         & Euler & Aitken & weighted-averages \\
        \hline
        \# of subintervals & 13 & 13 & 4\\
		Relative error & 1.4131 $\times\;10^{-6}$ & 1.1921 $\times\;10^{-6}$ & 1.4058 $\times\;10^{-6}$\\
        \hline
    \end{tabular}
    \label{T.p2}
\end{center}
\end{table}

\begin{table}[!htbp]
\begin{center}
\renewcommand{\arraystretch}{1.2}
\setlength{\tabcolsep}{12pt}
\caption{Comparison of the three extrapolation methods for Case 3.}
    \begin{tabular}{|c|c|c|c|}
        \hline
         & Euler & Aitken & weighted-averages \\
        \hline
        \# of subintervals & 13 & 13 & 5\\
		Relative error & 4.9227 $\times\;10^{-7}$ & 4.9227 $\times\;10^{-7}$ & 4.9799 $\times\;10^{-7}$\\
        \hline
    \end{tabular}
    \label{T.p3}
\end{center}
\end{table}

\bibliographystyle{plainnat}
\bibliography{potentialrefs}

\end{document}